\documentclass{siamltex}

\usepackage{graphicx}
\usepackage{amsmath}
\usepackage{amsfonts}
\usepackage{amssymb}
\usepackage{amsthm}
\usepackage{newlfont}
\usepackage{lscape}
\usepackage{graphicx}
\usepackage{multirow}

\pdfoutput=1

\begin{document}

\title{Calibration of Chaotic Models for Interest Rates}
\author{M. R.  Grasselli\thanks{Department of Mathematics and Statistics, McMaster University, Hamilton L8S 4K1, ON, Canada. Supported by the Natural Sciences and Engineering Research Council of Canada and MITACS.} \and T. Tsujimoto \thanks{School of Economics and Finance, University of St Andrews, Fife KY16 9AL, Scotland, UK. Supported by the Scottish Institute for
Research in Economics.}}
\date{December 01, 2010}
\maketitle

\begin{abstract}In this paper we calibrate chaotic models for interest rates to market data using a polynomial--exponential parametrization for the chaos coefficients. We identify a subclass of one--variable models that allow us to introduce complexity from higher order chaos in a controlled way while retaining considerable analytic tractability. In particular we derive explicit expressions for bond and option prices in a one--variable third chaos model in terms of elementary combinations of normal density and cumulative distribution functions. We then compare the calibration performance of chaos models with that of well--known benchmark models. For term structure calibration we find that chaos models are comparable to the Svensson model, with the advantage of guaranteed positivity and consistency with a dynamic stochastic evolution of interest rates. For calibration to option data, chaos models outperform the Hull and White and rational lognormal models and are comparable to LIBOR market models.    
\end{abstract}




\section{Introduction}

The purpose of this paper is to investigate the calibration performance of interest rate models based on the Wiener chaos expansion. Chaotic models were introduced in \cite{HughstonRafailidis05} as an axiomatic framework for interest rates satisfying both no arbitrage and positivity conditions, following the line of research initiated by \cite{FlesakerHughston96}, where zero--coupon bond prices are modeled as
\begin{equation}
\label{bondM}
P_{tT}=\frac{\int_{T}^\infty M_{ts}ds}{\int_{t}^\infty M_{ts}ds},
\end{equation}
for a parametrized family of positive martingales $M_{ts}$. It was then shown in \cite{Rogers97} and \cite{Rutkowski97} that one can focus instead on modeling a supermartingale $V_t$ which is related to the martingales $M_{ts}$ through
\begin{equation}
\label{FH}
V_t=\int_t^\infty M_{ts}ds,
\end{equation}
The chaotic approach is derived from the observation that $V_t$ can itself be written as the conditional variance of a terminal random variable $X_{\infty}$. This square integrable random variable then has a unique Wiener chaos decomposition, and studying the way its expansion coefficients affect the corresponding interest rates model becomes the subject of the theory.

Notice the successive simplifications in the class of objects that one needs to model: from an entire family of martingales $M_{ts}$ in the Flesaker--Hughston framework, to the process $V_t$ in the potential approach, and finally the single random variable $X_{\infty}$ in chaotic models. Arguably, each step of the way reduces the arbitrariness in the modeling exercise, with presumed advantages for calibration to real data. In particular, the chaos expansion allows one to successively introduce randomness in the model in a way that can, in principle, be made to correspond to the increased complexity of financial instruments under consideration. Such are the types of statements that we propose to put to test in this paper.

Models in the Flesaker--Hughston framework have been implemented and calibrated, for example, in \cite{Cairns04b} and \cite{Goldberg98}, whereas  implementations of the potential approach can be found in  \cite{KlugeRogers08} and \cite{RogersZane97}. To our knowledge, we present here the first practical  implementation and calibration to market data of chaotic interest rate models. We adopt the day-by-day calibration methodology used in \cite{GatarekBachertMaksymiuk06}, \cite{JarrowLiZhao07}, and \cite{RapisardaSilvotti01}. The motivation for this is to capture the prices of liquid interest rate derivatives such as caps and swaptions by a model as parsimonious as possible, which can then be used for pricing and hedging of exotic options, such as the Chooser flexible cap and Bermudan Swaption. 

After reviewing the chaotic approach in Section \ref{approach}, we implement to two separate calibration exercises. In Section \ref{term_calibration} we consider the calibration of chaotic models to the observed term structure of interest rates. For comparison, we use the Nelson--Siegel \cite{NelsonSiegel87} and 
Svensson \cite{Svensson95} models for forward rates as benchmarks. These so called descriptive models \cite{Cairns98} are examples of a general exponential--polynomial class of models analyzed in \cite{BjorkChristensen99}, which we use as motivation for the parametric form we adopt for the chaos coefficient functions. We find in this section that chaotic models perform comparably to descriptive models with the same number of parameters, with the advantage of avoiding problems with positivity and consistency.

We then move to a full calibration to yields and option prices in Section \ref{option}. We recall known expressions for option prices in a second chaos model, derive the corresponding formulas in a third chaos model, and calibrate them to two separate data sets in three different ways. For comparison, we also calibrate the Hull and White model \cite{HullWhite90}, the rational lognormal proposed in \cite{FlesakerHughston96} and a lognormal LIBOR market model \cite{GatarekBachertMaksymiuk06}. We find that chaos models generally perform much better than the Hull and White and rational lognormal models and have fitting errors comparable to those obtained with a LIBOR model. When we apply an information criterion that takes into account the different number of parameters in the models, we find that one of our third chaos models consistently outperforms the LIBOR market model. Section \ref{conclusion} concludes the paper by summarizing the results and pointing to future research in the area.
 
%

\section{The chaotic approach}
\label{approach}

We review the framework proposed in  \cite{HughstonRafailidis05}, whereby a general positive interest rate model with no arbitrage opportunities is associated with a square integrable terminal random variable $X_\infty$, which can then be modeled using a Wiener chaos expansion. 

\subsection{Basic definitions}
\label{definitions}

 Let $(\Omega,{\cal F},P)$ be a probability space equipped with the standard augmented filtration $({\cal F}_t)_{0\leq t <\infty}$ generated by a $k$--dimensional Brownian motion 
$W_t$ and suppose that $X_\infty \in L^2(\Omega,{\cal F},P)$ is a random variable with the property that $X_\infty$ is not ${\cal F}_t$--measurable for any finite value of $t$ and $E[X_\infty]=0$. Using the martingale representation theorem, we can write 
\begin{equation}
\label{sigma}
X_\infty =\int_0^\infty \sigma_s dW_s
\end{equation}
for an adapted process $\sigma_t$. Denoting the conditional variance of $X_\infty$ with respect to the $\sigma$--algebra ${\cal F}_t$ by
\begin{equation}
\label{cond_var}
V_t=E\left[\left.(X_\infty-E_t[X_\infty])^2\right|{\cal F}_t\right],  
\end{equation}
it follows from the It\^{o} isometry that 
\begin{equation}
V_t=E\left[\left.\left(\int_t^\infty \sigma_s dW_s\right)^{2}\right|{\cal F}_t\right]=E\left[\left.\int_t^\infty \sigma_s^2 ds\right|{\cal F}_t\right].
\end{equation}
Defining the increasing adapted process
\begin{equation}
A_t=\int_0^t \sigma^2_s ds.
\end{equation}
we have that 
\begin{equation}
\label{doob}
V_t=E[A_\infty-A_t|{\cal F}_t],
\end{equation}
from which it follows that $V_t$ is a potential, that is, a positive supermartingale satisfying $E[V_{t}]\to 0$ as $t\to\infty$. It is then well--known (see, for example, \cite{Rogers97}) that $V_{t}$ can be used as a state--price density for an arbitrage--free interest rate model in which the price of a zero--coupon bond with maturity $T$ is given by 
\begin{equation}
\label{bond}
P_{tT} =\frac{E\left[V_T|{\cal F}_t\right]}{V_t}=\frac{Z_{tT}}{Z_{tt}}, \qquad 0\leq t \leq T <\infty,
\end{equation}
where 
\begin{equation}
Z_{tT}=E[V_T|{\cal F}_t],  \qquad 0\leq t \leq T <\infty.
\end{equation}
It follows from \eqref{doob} that for each fixed $t$ the processes $Z_{tT}$, and consequently the bond prices $P_{tT}$, are decreasing functions of the maturity $T$, which in turn implies that all instantaneous forward rates 
\begin{equation}
f_{tT}=-\frac{\partial}{\partial T}\log P_{tT}
\end{equation}
are automatically positive in this framework. Furthermore, the short--rate $r_t$ and the market price of risk vector $\lambda_t$ are determined by the dynamics of $V_t$ as follows:
\begin{equation}
\label{Vsde}
dV_t=-r_t V_t dt- \lambda_t V_tdW_t, \qquad V_0>0.
\end{equation} 

For later use, we introduced the family of positive martingales 
\begin{equation}
M_{ts}=E[\sigma^2_s|{\cal F}_t], \qquad 0\leq t \leq s <\infty,
\end{equation}
so that $V_t$ can be written as in \eqref{FH} and bond prices as in \eqref{bondM}.

Returning to the squared integrable random variable $X_\infty$, the Wiener chaos decomposition (see \cite{Nualart96} and \cite{Wiener38}) says that it can be written as the 
$L^2$--convergent sum
\begin{align}
X_\infty =&
\int^\infty_0\phi_1(s)dW_s
+\int^\infty_0\int^s_0\phi_2(s,s_1) dW_{s_1}dW_s  \nonumber \\
&+\int^\infty_0\int^s_0\int^{s_1}_0\phi_3(s,s_1,s_2)dW_{s_2}dW_{s_1}dW_s+\cdots,
\end{align}
for deterministic square--integrable functions $\phi_1(s),\phi_2(s,s_1),\phi_3(s,s_1,s_2), \ldots$ called the chaos coefficients of $X_\infty$. We say that an interest rate model is an $n$-th chaos model if the decomposition for the random variable $X_\infty$ can be completely determined by its first $n$ coefficient functions.  

The basic insight of the chaotic approach is that the decomposition above provides a way to add complexity to an interest rate model in a controlled manner. For example, as we will see in the next section, the first order chaos correspond to deterministic interest rate models, whereas the second order chaos give rise to stochastic interest rate models with randomness governed by a parametric family of Gaussian processes. 

More importantly for calibration purposes, the increased complexity in the interest models should be related to the instruments that are available in the market. In what follows, we propose a systematic way to calibrate chaotic interest models starting with bond prices alone and then gradually increasing the complexity of market instruments included in the calibration. Our general strategy will consist of choosing general parametric forms for the deterministic functions $\phi_i$, $i=1,2,\ldots$ and then fit the parameters to market data according to the bond and option pricing formulas emerging from the models. 

\subsection{Option Pricing}
\label{prices}

We now describe how the prices of the most common interest rate options can be written in terms of the quantities defined in Section \ref{definitions}. Since 
$V_t$ is a state--price density, it follows that the price at time $t$ of a derivative with payoff $H_T$ is given by 
\begin{equation}
\label{derivatives}
H_t=\frac{E[V_TH_T|{\cal F}_t]}{V_t}.
\end{equation}
We see that expression \eqref{bond} for bond prices is simply a special case of \eqref{derivatives} with $H_T=P_{TT}=1$. We apply this general expression to European put options, caplets and swaptions, since these will be the only ones needed in our calibration section, although similarly results also hold for other interest rate derivatives, notably call options and floors (see \cite{Tsujimoto10}). We follow the standard definitions and notations in \cite{BrigoMercurio06}.

A put option with maturity $t$ and strike price $K$ written on a bond with maturity $T\geq t$ correspond to the following payoff
\begin{equation}
(K-P_{tT})^+.
\end{equation}
Using \eqref{derivatives}, we see that the price of the a bond option at time $s\leq t\leq T$ is given by
\begin{equation}
\label{put}
p(s,t,T,K)=\frac{E[V_t(K-P_{tT})^+|{\cal F}_{s}]}{V_s}=\frac{E[(KZ_{tt}-Z_{tT})^+|{\cal F}_{s}]}{V_s}.
\end{equation}


A caplet is a call option with strike $K$ and maturity $T$ written on a spot LIBOR rate $L(t,T)$ for the time interval $[t,T]$, that is, it is defined by the payoff 
\begin{equation}
\label{caplet_pay}
{\cal N}(T-t)(L(t,T)-K)^+,
\end{equation}
where ${\cal N}$ is a fixed notional amount. Recalling that the spot LIBOR rate is defined as
\begin{equation}
L(t,T):=\frac{1}{T-t}\left(\frac{1}{P_{tT}}-1\right)=\frac{1}{T-t}\left(\frac{V_{t}}{Z_{tT}}-1\right)
\end{equation}
we see from \eqref{derivatives} that the value of a caplet at time $s\leq t$ is
\begin{equation}
\label{caplet}
\begin{aligned}
\text{Cpl}(s,t,T,{\cal N},K)
=&\frac{E\left[V_{T}{\cal N}(T-t)(L(t,T)-K)^+|{\cal F}_s\right]}{V_s}\\
=&\frac{{\cal N}}{V_s}E\left[\left.\big(Z_{tt}-Z_{tT}(1+K(T-t))\big)^+\right|{\cal F}_s\right]\\
\end{aligned}
\end{equation}
Comparing \eqref{caplet} and \eqref{put} it is easy to deduce the following well--known relation \cite[page 41]{BrigoMercurio06}: 
\begin{equation}
\label{caplet_put}
\text{Cpl}(s,t,T,{\cal N},K)={\cal N}(1+K(T-t))p\left(s,t,T,\frac{1}{1+K(T-t)}\right),
\end{equation}
which allows us to obtain prices for caplets provided we know how to calculate the prices of bond put options. 

Observe that the payoff \eqref{caplet_pay} can be rewritten as 
\begin{equation}
{\cal N}(T-t)(F(t,t,T)-K)^+,
\end{equation}
where the forward LIBOR rate $F(s,t,T)$ is defined as
\begin{equation}
\label{forward_libor}
F(s,t,T):=\frac{1}{T-t}\left(\frac{P_{st}}{P_{sT}}-1\right) ,\qquad 0\leq s\leq t\leq T <\infty.
\end{equation}
Using this, we can define the caplet implied volatility as the value $v_{t,T}$ satisfying 
\begin{equation}
\text{Cpl}(0,t,T,{\cal N},K)={\cal N}P_{0T}(T-t)\text{Black}\left(K,F(0,t,T),v_{t,T}\sqrt{t}\right),
\end{equation}
where 
\begin{equation}
\label{black}
\text{Black}(K,F,v):=F\Phi\left(\frac{\log(F/K)+\frac{v^2}{2}}{v}\right)-K\Phi\left(\frac{\log(F/K)-\frac{v^2}{2}}{v}\right),
\end{equation}
is the Black formula and $\Phi(\cdot)$ denotes the standard normal cumulative distribution function. We say that a caplet is at-the-money (ATM) at time $t=0$ if 
$K=F(0,t,T)$. Given a set of maturities $T_i$, the so--called  term structure of caplet volatility is the implied volatility curve obtained from ATM caplets of consecutive maturities, that is, a function of the form  $T_i\rightarrow v_{T_{i-1},T_i}$. 

A payer Swaption with maturity $t$ is defined by the payoff 
\begin{equation}
\label{swaption}
\left({\cal N}\sum_{i=1}^n (T_{i}-T_{i-1})P_{t T_i}(F(t,T_{i-1},T_i)-K)\right)^+,
\end{equation}
where ${\cal T}=(T_0,T_1,\ldots,T_n)$ is a set of future dates with $T_0=t$ and ${\cal N}$ is the notional. The total length of the time interval $(T_n-T_0)$ is called the tenor of the swaption. 

Recalling the definition of the forward LIBOR rate in \eqref{forward_libor}, we obtain from \eqref{derivatives} that the value of the payer swaption at time $s\leq t$ is given by 
\begin{align}
\text{Swp}(s,{\cal T},{\cal N},K)
=&\frac{1}{V_s}E\left[\left.V_t\left({\cal N}\sum_{i=1}^n (T_{i}-T_{i-1})P_{t T_i}(F(t,T_{i-1},T_i)-K)\right)^+\right|{\cal F}_t\right] \nonumber \\
=&\frac{{\cal N}}{V_s}E\left[\left.\left(Z_{tt}-Z_{tT_n}-K\sum_{i=1}^n (T_i-T_{i-1})Z_{tT_i}\right)^+\right|{\cal F}_t\right].\label{swaption_formula}
\end{align}
Observe that the payoff in \eqref{swaption} can be rewritten as 
\begin{equation}
\label{swaption2}
\left({\cal N}\sum_{i=1}^n (T_{i}-T_{i-1})P_{t T_i}(S(t,t,T_n)-K)\right)^+,
\end{equation}
where the forward swap rate $S(s,t,T_n)$ is defined as
\begin{equation}
S(s,t,T_n):=\frac{P_{st}-P_{sT_n}}{\sum_{i=1}^n (T_i-T_{i-1})P_{sT_i}}, \quad 0\leq s\leq t\leq T_n<\infty.
\end{equation}
Similarly to caplets, we can use this to define the swaption implied volatility as the value $v_{t,T_n}$ satisfying 
\begin{equation}
\text{Swp}(0,{\cal T},{\cal N},K)={\cal N}\sum_{i=1}^n(T_i-T_{i-1})P_{0T_i}\text{Black}\left(K,S(0,t,T_n),v_{t,T_n}\sqrt{t}\right).
\end{equation}
The swaption is said to be at-the-money (ATM) at time $t=0$ if $K=S(0,t,T_n)$.

This concludes our general overview of chaotic models. We will have a chance to explore these formulas in more detail after we introduce specific models of different chaos orders in the calibration sections that come next. For simplicity, we will assume from now on that the underlying Brownian motion $W_t$ is one dimensional, although extensions to the multidimensional case should be relatively straightforward.

\section{Term structure calibration}
\label{term_calibration}

We first consider the calibration of the chaos models to the bond prices $P_{0T}$ observed at time $t=0$, or equivalently, to the corresponding forward rates 
\[f_{0T}=-\frac{\partial}{\partial T}\log P_{0T}.\]

Historically, the problem of characterizing the term structure of interest rates implied by bond prices at time $t=0$ has been extensively tackle by so--called descriptive models, which consists of choosing a parametric form for the forward rates $f_{0T}$ and then fit the relatively small number of parameters to the observed yield curve. Popular examples of this approach are:
\begin{equation}
\label{descriptive}
f_{0T}=\begin{cases}
&b_0+[b_1+b_2T]e^{-c_1T}\quad \text{(Nelson and Siegel)} \\
&b_0+[b_1+b_2T]e^{-c_1T}+b_3Te^{-c_2T}\quad \text{(Svensson)} \\
&b_0+\sum_{i=1}^4b_ie^{-c_iT}\quad \text{(Cairns)},
\end{cases}
\end{equation}
introduced respectively in \cite{NelsonSiegel87}, \cite{Svensson95} and \cite{Cairns98}. All of these examples fall within the exponential-polynomial family of forward rates analyzed in \cite{BjorkChristensen99}:  
\begin{equation}
f_{0T}=L_0(T)+\sum_{i=1}^nL_i(T)e^{-c_iT}\quad \text{where}\quad L_i(T)=\sum_{j=0}^{k_i}b_{ij}T^j.
\end{equation}

As explained in \cite{Cairns98}, descriptive models provide a snapshot, but not a movie. That is to say, they do not provide a model for the stochastic evolution of interest rates. Nevertheless, they can be used to price simple instruments such as forward rate agreements and swaps, which do not depend on the dynamics of interest rates. They have also been widely used by central banks for the purposes of monetary policy (see the table presented in \cite{Filipovic00}). 

For more complicated interest rate derivatives, it becomes necessary to embed such static descriptions into a fully dynamic model. For example, one can use the fitted forward rate curve $f_{0T}$ as the initial term structure for a model within the Heath--Jarrow--Morton framework \cite{HeathJarrowMorton92}, supplementing it with the specification of the volatility structure. In other words, the specification of the initial term structure used as an input is done separately from its stochastic evolution.

By contrast, instead of focusing on the forward rates $f_{0T}$, we propose to use the observed term structure at time $t=0$ to calibrate the deterministic functions 
$\phi_i(T)$, which then completely specify the stochastic evolution of interest rates. Since these are markedly different procedures, their performance need to be compared using real observed data, as we do later in this section. Before that we need to describe the term structure of interest rates for chaotic models of different orders in mode detail.

\subsection{Term structure in first chaos models}

In a first chaos model we have 
\begin{equation}
X_\infty =
\int^\infty_0\phi_1(s)dW_s
\end{equation}
where $\phi_1(s)$ is a deterministic function of one variable. Comparing this with \eqref{sigma} we conclude that $\sigma_s\equiv \phi_1(s)$, so that 
\begin{equation}
\label{Vfirst}
V_t=E\left[\left.\int_t^\infty \sigma_s^2 ds\right|{\cal F}_t\right]=\int_t^\infty \phi_1(s)^2ds,
\end{equation}
which implies that
\[dV_t=-\phi_1(t)^2 dt.\]
Comparing this with \eqref{Vsde} then gives
\begin{align}
r_t &=\frac{\phi_1^2(t)}{V_t}=\frac{\phi_1^2(t)}{\int_t^\infty \phi_1^2(s)ds} \\
\lambda_t & = 0.
\end{align}
In other words, in first chaos models the state--price density is a deterministic function, implying that the interest rates associated with it are themselves deterministic. In particular, $M_{ts}=\phi_1^2(s)$ constitutes a family of constant martingales for each value of $s$ and bond prices are given by 
\begin{equation}
\label{Pfirst}
P_{tT}=\frac{E\left[V_T|{\cal F}_t\right]}{V_t}=\frac{\int_{T}^\infty M_{ts}ds}{\int_{t}^\infty M_{ts}ds}=\frac{\int_T^\infty \phi_1^2(s)ds}{\int_t^\infty \phi_1^2(s)ds}=e^{-\int_t^T r_s ds}.
\end{equation}
It follows that instantaneous forward rates in first chaos models have the form 
\begin{equation}
\label{forward_phi}
 f_{tT}=-\frac{\partial}{\partial T}\log P_{tT}=\frac{\phi_1^2(T)}{\int_T^\infty \phi_1(s)^2 ds}=\frac{\phi_1^2(T)}{V_T}=r_T,
\end{equation}
which are manifestly positive, as they should be in any chaotic model. Moreover, because they represent market expectations at time $t$ for what the (deterministic) short--rate will be at time $T$, instantaneous forward rates in first chaos models are independent of $t$ and always equal to $r_T$.

As already noticed in \cite{HughstonRafailidis05}, these observations show that first chaos models characterize general arbitrage--free, positive, deterministic interest rate term structures, which should not be immediately disregarded as trivial, since they constitute the starting point for the majority of the applications of interest rate theory. Chaotic models then provide a well--defined way to embed such deterministic models into their stochastic generalizations, as we will demonstrate in the remainder of this paper.

From the point of view of calibration, we can calibrate a first chaos model to the term structure at time $t=0$ by focusing on the equation
\begin{equation}
P_{0T}=\frac{\int_T^\infty \phi_1(s)ds}{\int_0^\infty \phi_1(s)ds}.
\end{equation}
Because of the deterministic nature of first chaos models, this is enough to characterize bond prices at all times, since it follows from 
\eqref{Pfirst} that $P_{tT}=P_{0T}/P_{0t}$. 

\subsection{Term structure in second chaos models}

In a second chaos model we have 
\[X_\infty = \int^\infty_0\phi_1(s)dW_s+\int^\infty_0\int^s_0\phi_2(s,u) dW_{u}dW_s,\]
for deterministic functions $\phi_1(s)$ and $\phi_2(s,u)$.  Comparing this with \eqref{sigma} gives
\begin{equation}
\sigma_t=\phi_1(t)+\int_0^t\phi_2(s,u)dW_{u}.
\end{equation}
It then follows from the conditional Ito isometry that
\begin{align}
M_{ts}&=E[\sigma^2_s|{\cal F}_t] \nonumber \\
&=\left(\phi_1(s)+\int_0^t \phi_2(s,u)dW_{u}\right)^2+\int_t^s\phi_2^2(s,u)du \nonumber \\
&=R^2_{ts}-Q_{ts}+Q_{ss}\label{Mts}
\end{align}
where, for each fixed value $s\in[t,\infty)$, the process
\begin{equation}
R_{ts}:=\phi_1(s)+\int_0^t \phi_2(s,u)dW_{u}
\end{equation}
is a martingale in the variable $t$ with quadratic variation 
\begin{equation}
Q_{ts}:=\int_0^t\phi_2^2(s,u)du.
\end{equation}
  
Setting $t=0$ in \eqref{Mts} and using \eqref{bondM} we have that the initial term structure of bond prices in second chaos models is generically given by 
\begin{equation}
\label{initial_second}
P_{0T}=\frac{\int_{T}^\infty \left(\phi_1^2(s)+\int_t^s\phi^2_2(s,u)du\right)ds}
{\int_{0}^\infty \left(\phi_1^2(s)+\int_0^s\phi^2_2(s,u)du\right)ds}.
\end{equation}

Expression \eqref{initial_second} looks uncomfortably complicated for the purpose of calibrating a relatively simple object such as $P_{0T}$. To achieve greater tractability, we now recall the class of {\em factorizable} second chaos models introduced in \cite{HughstonRafailidis05}, whereby 
\begin{equation}
\phi_1(s)=\alpha(s),\qquad \phi_2(s,u)=\beta(s)\gamma(u).
\end{equation}
for deterministic functions $\alpha$, $\beta$ and $\gamma$. In other words, factorizable second chaos models correspond to random variables $X_\infty$ of the form
\begin{equation}
X_{\infty}
=\int^{\infty}_0\alpha(s)dW_{s}+\int^{\infty}_0\int^{s}_0\beta(s)\gamma(u)dW_{u}dW_{s}.
\end{equation}
for $0\leq u \leq s<\infty$. In this case we have that  
\begin{equation}
\label{Rts}
R_{ts}=\alpha(s) +\beta(s)R_t, \qquad
Q_{ts}=\beta^2(s)Q_t,
\end{equation}
where 
\begin{equation}
R_t:=\int^t_0\gamma(u) dW_{u}
\end{equation}
is a Gaussian martingale with quadratic variation $Q_t:=\int_0^t \gamma^2(u)du$. Inserting \eqref{Rts} into \eqref{Mts} we obtain
\begin{equation}
\label{Z_fact}
\begin{aligned}
Z_{tT}&=\int_T^\infty M_{ts}ds \\
&=\int^\infty_T\left[\left(\alpha(s)+\beta(s)R_t\right)^2-\beta^2_sQ_t+\beta^2(s)Q_s\right]ds\\
&=A(T)+B(T) R_t+C(T)(R^2_t-Q_t)
\end{aligned}
\end{equation}
where 
\begin{equation}
\label{ABC}
\begin{aligned}
A(T)&=\int^{\infty}_T[\alpha^2(s)+\beta^2(s)Q_s]ds, \\
B(T)&=\int^{\infty}_T 2\alpha(s)\beta(s)ds, \\
C(T)&=\int^{\infty}_T\beta^2(s)ds.
\end{aligned}
\end{equation}

Using the fact that $V_t=Z_{tt}$, it is easy to see that the state--price density in this case is given by 
\begin{equation}
V_t=A(t)+B(t)R_t+C(t)(R^2_t-Q_t),
\end{equation} 
which is a simple quadratic polynomial in the Gaussian process $R_t$. Therefore bond prices in a factorizable second chaos model are given by the following ration of quadratic polynomials in $R_t$:
\begin{equation}
\label{bond_fac}
P_{tT}=\frac{A(T)+B(T) R_t+C(T)(R^2_t-Q_t)}
{A(t)+B(t)R_t+C(t)(R^2_t-Q_t)}.
\end{equation}
In particular, the initial term structure is given by 
\begin{equation}
\label{initial_fac}
P_{0T}=\frac{A(T)}{A(0)}=\frac{\int^{\infty}_T[\alpha^2(s)+\beta^2(s)Q_s]ds}{\int^{\infty}_0[\alpha^2(s)+\beta^2(s)Q_s]ds}.
\end{equation}

Despite being much simpler than \eqref{initial_second}, we observe that \eqref{initial_fac} still offers too much freedom for the choice of parametric forms for the deterministic functions involved. To make this point more explicitly, observe that from the strict point of view of calibrating the initial term structure, we see that a factorizable second chaos model behaves exactly like a first chaos model if we identify 
\begin{equation}
\label{int_gamma}
\phi_1(s)=\alpha^2(s)+\beta^2(s)\int_0^s \gamma^2(u)du,
\end{equation}
for arbitrary choices of $\alpha,\beta$ and $\gamma$. This is to be expected in all chaos models, since the initial bond prices available for calibration correspond to the one--variable function $P_{0T}$, regardless of how many deterministic functions $\phi_i$ we wish to calibrate. Of course, once calibrated, chaos models of different orders give rise to very different {\em future} term structures: a deterministic one with $P_{tT}=P_{0T}/P_{0t}$ for first chaos models and a fully stochastic one where $P_{tT}$ is given by \eqref{bond_fac} for factorizable second chaos models, for example. 

These remarks point to a general yet delicate feature of chaotic models: one can achieve increased stochastic complexity in the form of higher order chaos at the expense of having to calibrate complicated expressions that are not necessarily supported by the data available. Our approach to deal with this problem consists of increasing the complexity of the models as gradually as possible, while carefully evaluating their statistical performance when calibrated to real data.

Accordingly, we propose to investigate the subclass of factorizable second chaos models with $\gamma(u)\equiv 1$ and $\beta(s)$ satisfying 
\begin{equation}
\label{beta}
\int_0^\infty \beta(s)^2sds <\infty,
\end{equation}
that is, models corresponding to the terminal random variable
\begin{equation}
X_{\infty}
=\int^{\infty}_0\!\!\!\alpha(s)dW_{s}+\int^{\infty}_0\!\!\!\beta(s)\int^{s}_0dW_{u}dW_{s}=
\int^{\infty}_0\!\!\!\alpha(s)dW_{s}+\int^{\infty}_0\!\!\!\beta(s)W_sdW_{s}.
\end{equation}
As we can see, these {\em one--variable} second chaos models provide the simplest stochastic extensions of first order chaos, for which we have  
that $R_t=W_t$ and the initial term structure of bond prices is given by 
\begin{equation}
\label{initial_one_sec}
P_{0T}=\frac{\int^{\infty}_T[\alpha^2(s)+\beta^2(s)s]ds}{\int^{\infty}_0[\alpha^2(s)+\beta^2(s)s]ds}.
\end{equation}

\subsection{Term structure in higher order chaos models}
\label{higher}

In a general chaos model we that
\begin{eqnarray*}
Z_{0T}&=&\int^\infty_T \mathbb{E}\left[\left(\phi_1(s_1) +\int^{s_1}_0\phi_2(s_1,s_2) dW_{s_2}+\cdots
\right)^2\right]ds_1 \\
&=&\int^\infty_T \left(\phi^2_1(s_1) +\int^{s_1}_0\phi^2_2(s_1,s_2) ds_2
+\cdots\right)ds_1.
\end{eqnarray*}
Therefore 
\begin{equation}
\label{initial_general}
P_{0T}=\frac{Z_{0T}}{Z_{0t}}=\frac{\int_T^\infty \psi(s)ds}{\int_t^\infty \psi(s)ds},
\end{equation}
where 
\begin{equation}
\label{psi_general}
\psi(s)=
\begin{cases}
 \phi^2_1(s) & \text{(first chaos)}\\
 \phi^2_1(s) +\int^{s}_0\phi^2_2(s,u) du & \text{(second chaos)}\\
 \phi^2_1(s) +\int^{s}_0\phi^2_2(s,u) du
+\int^{s}_0\int^{u}_0\phi^2_3(s,u,v)dvdu & \text{(third chaos)}\\
\quad\quad\vdots
\end{cases}
\end{equation}

In other words, as mentioned in the previous section, equation \eqref{initial_general} shows that from the strict point of view of calibrating the initial term structure 
$P_{0T}$, all chaos models behave like first order chaos models, albeit with a complicated choice of function $\psi(s)$. The differences between chaos models will become apparent, however, when we tackle the calibration to option prices in Section \ref{option}. For now, we find unnecessary to consider fully general higher order chaos and restrict ourselves to the simplest possible third chaos models. That is to say, we start with factorizable third chaos models of the form 
\begin{equation*}
X_\infty=\int^\infty_0\!\!\!\!\alpha(s)dW_s+\int^\infty_0\!\!\!\!\int^s_0\!\beta(s)\gamma(u)dW_{u}dW_s
+\int^\infty_0\!\!\!\!\int^s_0\!\!\!\int^{u}_0\!\!\delta(s)\epsilon(u)\zeta(v)dW_{v}dW_{u}dW_s,
\end{equation*}
for $0\leq v\leq u\leq s<\infty$. We then take $\gamma\equiv\epsilon\equiv\zeta\equiv 1$, $\beta$ satisfying \eqref{beta} and $\delta$ satisfying 
\begin{equation}
\label{gamma}
\int_0^\infty \delta(s)^2s^2ds <\infty
\end{equation}
to obtain the class of {\em one--variable} third chaos models of the form
\begin{equation}
\label{one_var_third}
\begin{aligned}
X_\infty&=\int^\infty_0\!\!\!\!\alpha(s)dW_s+\int^\infty_0\!\!\!\!\int^s_0\!\beta(s)dW_{u}dW_s
+\int^\infty_0\!\!\!\!\int^s_0\!\!\!\int^{u}_0\!\!\delta(s)dW_{v}dW_{u}dW_s \\
&=\int^\infty_0\left[\alpha(s)+\beta(s)W_s+\frac{1}{2}\delta(s)(W_s^2-s)\right]dW_s.
\end{aligned}
\end{equation}
We see from \eqref{one_var_third} that in this case
\begin{equation}
\sigma_s=\alpha(s)+\beta(s)W_s+\frac{1}{2}\delta(s)(W_s^2-s)
\end{equation}
Recalling that $M_{ts}=E[\sigma^2_s|{\cal F}_t]$, a direct calculation shows that
\begin{equation}
\label{Z_third}
\begin{aligned}
Z_{tT}=\int^\infty_T M_{ts}ds&=\widetilde A(T)+\widetilde B(T)W_t+\widetilde C(T)\left(W^2_t-t\right)\\
&+\widetilde D(T)\left(W^3_t-3tW_t\right)+\widetilde E(T)\left(W^4_t-6tW^2_t+3t^2\right),
\end{aligned}
\end{equation}
where the coefficients are
\begin{equation}
\begin{aligned}
\widetilde A(T)&=\int_T^\infty \left(\alpha^2(s)+s\beta^2(s)+\frac{s^2\delta(s)^2}{2}\right)ds \\
\widetilde B(T)&=\int_T^\infty 2\beta(s)(\alpha(s)+s\delta(s))ds \\
\widetilde C(T)&=\int_T^\infty \left(\beta^2(s)+\alpha(s)\delta(s)+\delta^2(s)s\right)ds \\
\widetilde D(T)&=\int_T^\infty \delta(s)\beta(s)ds \\
\widetilde E(T)&=\int_T^\infty \frac{\delta^2(s)}{4}ds
\end{aligned}
\end{equation}
Using again the fact that $V_t=Z_{tt}$, it follows that the state--price density is
\begin{equation}
\begin{aligned}
V_t&=\widetilde A(t)+\widetilde B(t)W_t+\widetilde C(t)\left(W^2_t\!-\!t\right)\\
&\widetilde D(t)\left(W^3_t\!-\!3tW_t\right)+\widetilde E(t)\left(W^4_t\!-\!6tW^2_t\!+\!3t^2\right).
\end{aligned}
\end{equation}
Therefore bond prices in a one--variable third chaos model are given by the following ratio of forth degree polynomials in $W_t$:
\begin{equation}
P_{tT}=\frac{\scriptstyle{\widetilde A(T)+\widetilde B(T)W_t+\widetilde C(T)\left(W^2_t-t\right)+\widetilde D(T)\left(W^3_t-3tW_t\right)+\widetilde E(T)\left(W^4_t-6tW^2_t+3t^2\right)}}{\scriptstyle{\widetilde A(t)+\widetilde B(t)W_t+\widetilde C(t)\left(W^2_t-t\right)+\widetilde D(t)\left(W^3_t-3tW_t\right)+\widetilde E(t)\left(W^4_t-6tW^2_t+3t^2\right)}}
\end{equation}
In particular, the initial term structure is given by
\begin{equation}
P_{0T}=\frac{\widetilde A(T)}{\widetilde A(0)}=\frac{\int_T^\infty \left(\alpha^2(s)+s\beta^2(s)+\frac{s^2\delta(s)^2}{2}\right)ds}{\int_0^\infty \left(\alpha^2(s)+s\beta^2(s)+\frac{s^2\delta(s)^2}{2}\right)ds}
\end{equation}

\subsection{Choices of the chaos coefficients}
\label{models}

Inspired by the parametric forms used in descriptive models, we propose to model the function $\phi_1$ itself in exponential--polynomial form. Observe that setting 
\begin{equation}
\label{parametric}
\phi_1(s)=\sum_{i=1}^nL_i(s)e^{-c_is},\quad \qquad L_i(s)=\sum_{j=0}^{k_i}b_{ij}s^j,
\end{equation}
where $L_0(s)=0$ and $c_i>0$ to guaranteed integrability, leads to a first chaos model in which forward rates are ratios of functions in exponential--polynomial form. We regard this {\em rational exponential--polynomial} family as a natural extension of the exponential--polynomial class considered in \cite{BjorkChristensen99}, with added flexibility and potentially better calibration performance. In particular, this class allows us to reproduce all empirical shapes commonly observed for forward rate curves, such as increasing, decreasing and hump-shaped functions. 

Having made this choice for the function $\phi_1$, equation \eqref{int_gamma} and its generalization \eqref{psi_general} naturally lead us to consider functions 
$\alpha,\beta,\gamma$ and $\delta$ all belonging to the exponential--polynomial class as well. 

\subsection{Calibration results}

We describe the details of the term structure calibration in Appendix \ref{term_procedure}. We calibrate 14 different chaos models using two distinct data set from the UK bond market: observations of bonds of different maturities at every other business day from January 1998 to January 1999 (a volatile market often exhibiting an inverted yield curve) and weekly observations from December 2002 to December 2005 (a more moderate market). For comparison, we also calibrate two of the descriptive models specified in \eqref{descriptive}.

We summarize our calibration results for the two data sets in Tables \ref{table:1} and \ref{table:2}. The first column in each table labels the models, starting with the descriptive Nelson--Siegel and Svensson models defined by the first two expression in \eqref{descriptive}, followed by the models \eqref{A1}--\eqref{A14} in Appendix \ref{term_procedure}. The second column characterizes the type of model, whereas the third one gives the number of calibrated parameters.  The remaining columns show the average values for the negative log--likelihood function, the Root-Mean-Squared Percentage Error and the Diebold-Mariano statistics with respect to the Svensson model, as described in Appendix \ref{term_procedure}. 
 
\begin{table}[h]
\caption{Term structure calibration for 1998-1999 (Volatile Market)}
\centering
\vspace{0.3cm}
\begin{tabular}{|c|l|c|r|c|r|}
\hline
 & Model & N  &  -L & RMSPE (\%) &  DM\\
\hline
Sv & Svensson & 6 & 160  & 0.70  & - \\
\hline
NS & Nelson--Siegel & 4 & 2101 & 2.67  & -4.45\\
\hline
1 & 1st chaos & 3 & 4420  & 4.44  & -11.46\\
\hline
2 & 1st chaos  & 5 & 250  & 0.86  & -3.54\\
\hline
3 & one-var 2nd chaos & 6 & 162  & 0.82  & -2.26\\
\hline
4 & one-var 2nd chaos  & 7 & 160  & 0.69  & 0.22\\
\hline
5 & one-var 2nd chaos  & 7 & 145  & 0.75  & -1.05\\
\hline
6 & factorizable 2nd chaos & 6 & 335  & 0.88  & -2.54\\
\hline
7 & factorizable 2nd chaos & 6 & 245  & 0.68  & 0.27\\
\hline
8 & factorizable 2nd chaos & 6 & 1245  & 1.26  & -3.81\\
\hline
9 & factorizable 2nd  chaos& 7 & 179  & 0.63  & 1.38\\
\hline
10 & factorizable 2nd  chaos & 7 & 153  & 0.72  & -1.07\\
\hline
11 & one-var 3rd chaos & 6 & 168  & 0.72  & -1.24\\
\hline
12 & one-var 3rd chaos  & 7 & 141  & 0.76  & -1.16\\
\hline
13 & one-var 3rd chaos  & 7 & 152  & 0.72 & -1.19\\
\hline
14 & one-var 3rd chaos  & 7 & 149  & 0.76  & -1.43\\
\hline
\end{tabular}  
\label{table:1}
\end{table}

\begin{table}[h]
\caption{Term structure calibration for 2002-2005 (Moderate Market)}
\centering
\vspace{0.3cm}
\begin{tabular}{|c|l|c|r|c|r|}
\hline
 & Model & N &  -L  & RMSPE (\%)  & DM\\
\hline
Sv & Svensson & 6 & 442  & 0.76  & -\\
\hline
NS & Nelson--Siegel  & 4 & 541 & 0.97  & -1.76\\
\hline
1 & 1st chaos  & 3 & 8716  & 3.96  & -3.50\\
\hline
2 & 1st chaos  & 5 & 438  & 0.99  & -1.99\\
\hline
3 & one-var 2nd chaos & 6 & 388  & 0.89  & -1.23\\
\hline
4 & one-var 2nd chaos  & 7 & 388  & 0.80  & -0.38\\
\hline
5 & one-var 2nd chaos  & 7 & 329  & 0.66  & 1.26\\
\hline
6 & factorizable 2nd chaos & 6 & 437  & 1.04  & -3.33\\
\hline
7 & factorizable 2nd chaos & 6 & 495  & 0.84  & -0.68\\
\hline
8 & factorizable 2nd chaos & 6 & 421  & 1.19  & -2.84\\
\hline
9 & factorizable 2nd  chaos & 7 & 365  & 0.82  & -0.78\\
\hline
10 & factorizable 2nd chaos  & 7 & 323  & 0.72  & 0.36\\
\hline
11 & one-var 3rd chaos & 6 & 388  & 0.87  & -1.06\\
\hline
12 & one-var 3rd chaos  & 7 & 350  & 0.78  & -0.11\\
\hline
13 & one-var 3rd chaos  & 7 & 367  & 0.68  & 1.24\\
\hline
14 & one-var 3rd chaos  & 7 & 325  & 0.69  & 0.60\\
\hline
\end{tabular}  
\label{table:2}
\end{table}
 
Starting with the results in Table \ref{table:1} for the volatile market, we first notice that, as expected, the Svensson model, which has 6 parameters, outperforms all three models that use a smaller number of parameters. On the other hand, out of five chaos models with 6 parameters, three are outperformed by the Svensson model, while the other two perform comparably. Finally, all chaos models with 7 parameters have a performance that is comparable to the Svensson model. We are led to conclude that chaos models do not offer a significant advantage with respect to the Svensson model under volatile market conditions, when yield curves can change from normal to inverted to humped shapes quiet rapidly. Moving on to Table \ref{table:2}, we observe  smaller RMSPEs and higher DM statistics uniformly across all chaotic models, but nevertheless not high enough to reject the null hypotheses that they have the same calibration performance as the Svensson model. 

More interestingly, both tables confirm our previous remarks in Section \ref{higher} that all chaos models behave similarly to a first chaos model from the point of view of calibrating the initial term structure. Indeed, we do not find significant difference in the calibration performances between chaos models of different orders. If anything, we notice that factorizable second chaos models perform slightly worse than their one--variable analogues, therefore justifying our assertion that one should avoid trying to calibrate the initial term structure using a complicated expression such as \eqref{initial_fac}.  

\begin{figure}[h]
\begin{center}
\includegraphics[width=\textwidth]{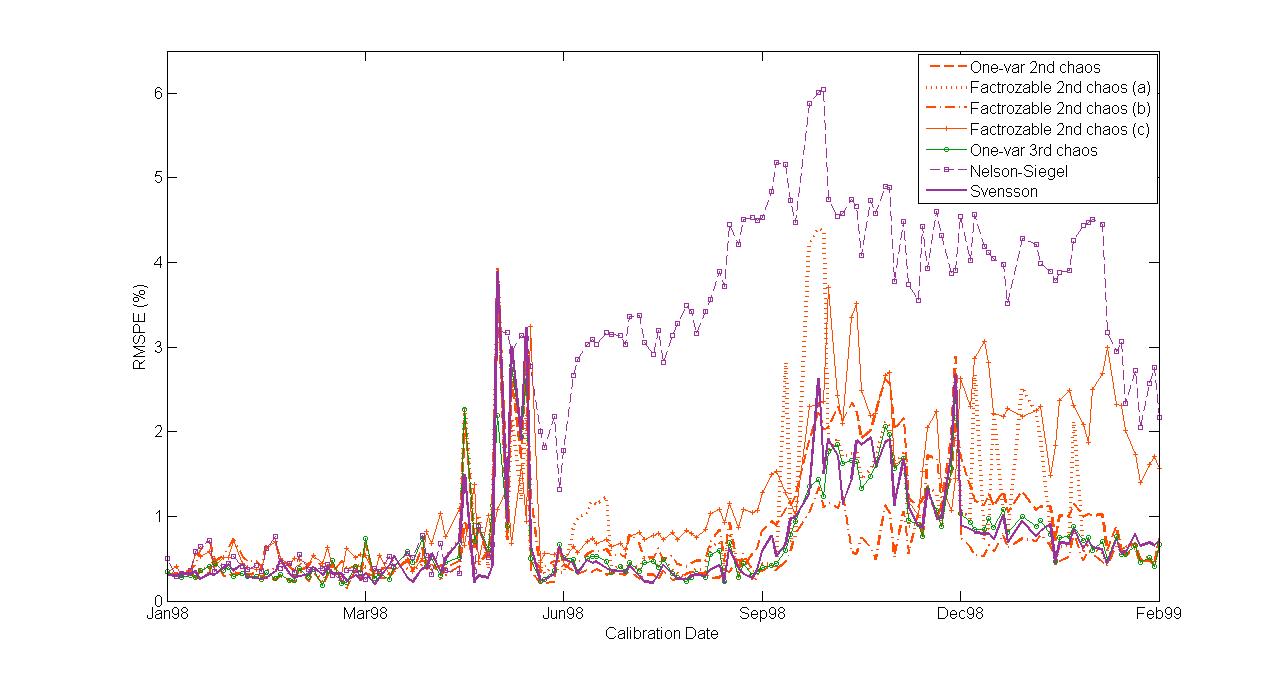}
\end{center}
\caption{Error for term structure calibration in 1998-1999.}\label{errors1}
\end{figure}

\begin{figure}[h]
\begin{center}
\includegraphics[width=\textwidth]{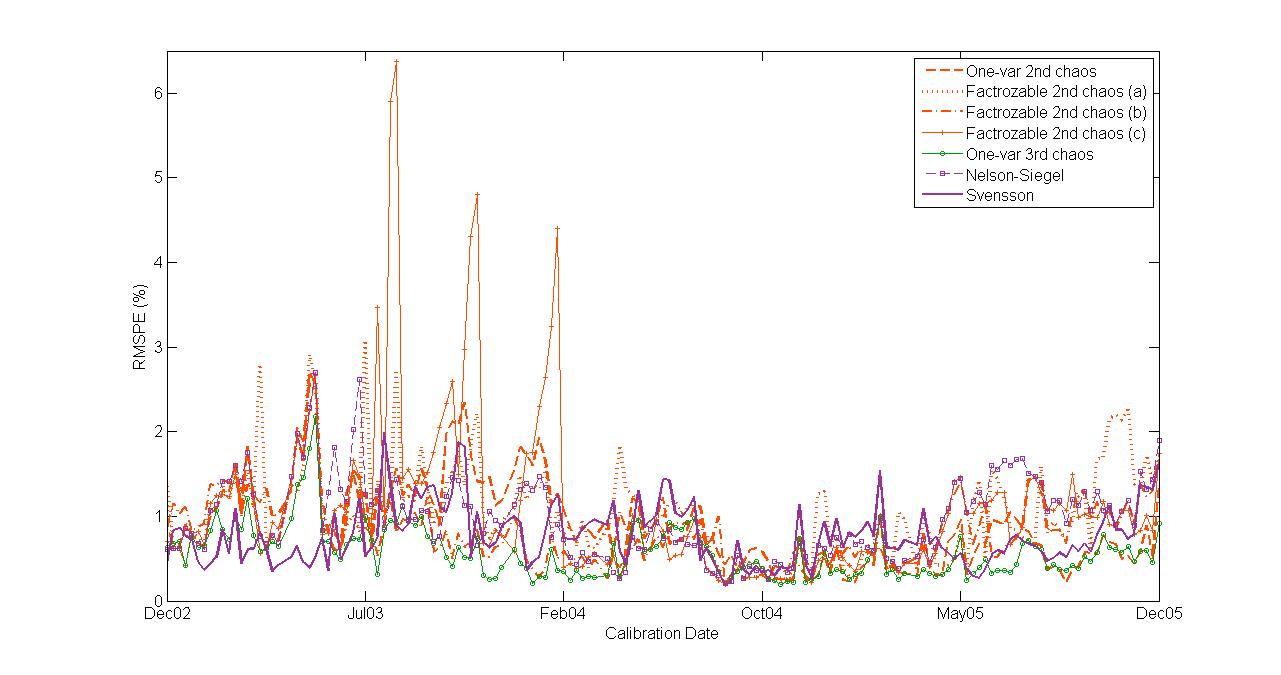}
\end{center}
\caption{Error for term structure calibration in 2002-2005.}\label{errors2}
\end{figure}

We illustrate the results by plotting the RMSPE as a function of time in Figures \ref{errors1} and \ref{errors2}. For comparison, we restrict ourselves to the chaotic models with 6 parameters in Tables \ref{table:1} and \ref{table:2}, as well as the Svensson and Nelson-Siegel models. 

The basic overall conclusion of this section is that, as far as the daily calibration of initial term structure is concerned, we cannot state that there exists a significant difference in the calibration performances between the Svensson and the chaotic models. However, as we mentioned before, chaotic models specify the stochastic evolution of interest rates, in addition to ensuring their positivity. In other words, the parameters calibrated from yield data may be applied directly to the volatility term structure. As we will see in the next section, this turns out to be greatly advantageous in the joint calibration of yields and option prices. 

\section{Option calibration}
\label{option}

In the previous section we showed that the chaotic and descriptive models have comparable performances when calibrated to yield curves. In this section, we take into account ATM European options, in particular caplets and swaptions. Since descriptive models for yield curves are silent regarding the future evolution of interest rates, we compare chaos models of different orders with other popular interest rate models in the literature, including short--rate models, models based on the potential approach and market models. But let us first analyze in greater detail the expressions for option prices in chaos models of different orders.

\subsection{Option prices in second chaos models}

Since in a first chaos model the price at time $0$ of an interest rate derivative with payoff $H_T$ reduces to the deterministic expression  
\begin{equation}
H_0=\frac{E[V_TH_T]}{V_0}=\frac{V_T}{V_0}H_T=e^{-\int_0^T r_s ds}H_T,
\end{equation}
any nontrivial option pricing should start with at least a second chaos model. We begin with bond put options in a factorizable second chaos, for which it has already been observed in \cite{HughstonRafailidis05} that \eqref{Z_fact} implies that 
\begin{equation}
KZ_{tt}-KZ_{tT}=\mathcal{P}^{(2)}_p(z),
\end{equation}
where $z=R_t/\sqrt{Q_t}$ is a standard normal random variable and
\begin{equation}
\mathcal{P}^{(2)}_p(z):=a_0+a_1 z+a_2 z^2
\end{equation} 
is a second degree polynomial with coefficients
\begin{equation}
\begin{aligned}
a_0&=-\big(A(T)-KA(t)\big)+\big(C(T)-KC(t)\big)Q_t \\
a_1&=-\big(B(T)-KB(t)\big)Q^{1/2}(t) \\
a_2&=-\big(C(T)-KC(t)\big)Q_t.
\end{aligned}
\end{equation}
Using \eqref{put}, it then follows that 
\begin{equation}
\label{put_fac_sec}
p(0,t,T,K)=\frac{1}{A(0)\sqrt{2\pi}}\int_{\mathcal{P}^{(2)}_p(z)\geq0}\mathcal{P}^{(2)}_p(z)e^{-\frac{z^2}{2}}dz.
\end{equation}
As shown in \cite{HughstonRafailidis05}, once we calculate the roots of $\mathcal{P}^{(2)}_p(z)$, we have that \eqref{put_fac_sec} reduces to a simple expression given in terms of the standard normal cumulative distribution function. Having calculated the price of a put option on a bond, we can easily find the price of a caplet using expression \eqref{caplet_put}.

Regarding the price of a payer swaption, it was also observed in \cite{HughstonRafailidis05} that 
\begin{equation}
Z_{tt}-Z_{tT_n}-K\sum_{i=1}^n (T_i-T_{i-1})Z_{tT_i}=\mathcal{P}^{(2)}_{Swp}(z)
\end{equation}
where $z$ is the standard normal random variable defined above and 
\begin{equation}
\mathcal{P}_{Swp}(z):=b_0+b_1 z+b_2 z^2
\end{equation} 
is a second degree polynomial with coefficients
\begin{equation}
\begin{aligned}
b_0=&\left(A(t)-A(T_n)-K\sum_{i=1}^n (T_i-T_{i-1})A(T_i)\right)\\
&-\left(C(t)-C(T_n)-K\sum_{i=1}^n (T_i-T_{i-1})C(T_i)\right)Q_t \\
b_1=&\left(B(t)-B(T_n)-K\sum_{i=1}^n (T_i-T_{i-1})B(T_i)\right)Q^{1/2}t \\
b_2=&\left(C(t)-C(T_n)-K\sum_{i=1}^n (T_i-T_{i-1})C(T_i)\right)Q_t.
\end{aligned}
\end{equation}
It then follows from \eqref{swaption_formula} that the price of a payer swaption in a factorizable second chaos model is
\begin{equation}
\label{swaption_fac_sec}
\text{Swp}(0,{\cal T},{\cal N},K)=\frac{1}{A(0)\sqrt{2\pi}}\int_{\mathcal{P}^{(2)}_{Swp}(z)\geq0}\mathcal{P}^{(2)}_{Swp}(z)e^{-\frac{z^2}{2}}dz.
\end{equation}
Once more, as shown in \cite{HughstonRafailidis05}, having calculated the roots of $\mathcal{P}^{(2)}_{Swp}(z)$, expression \eqref{swaption_fac_sec} reduces to an explicit formula in terms of the standard normal cumulative distribution and density functions. 

\subsection{Option prices in third chaos models}

We focus on one--variable third chaos models of the form \eqref{one_var_third}. Recalling expression \eqref{Z_third} for the process $Z_{tT}$ we have that
\begin{equation}
KZ_{tt}-Z_{tT}={\cal P}^{(4)}_p(z)
\end{equation}
where $z=W_t/\sqrt{t}$ is a standard normal random variable and 
\begin{equation}
\mathcal{P}^{(4)}_{c}(z):=c_0+c_1 z+c_2 z^2+c_3 z^3 + c_4 z^4
\end{equation} 
is a forth degree polynomial with coefficients
\begin{equation}
\begin{aligned}
c_0&=-\left(\widetilde A(T)-K\widetilde A(t)\right)+\left(\widetilde C(T)-K\widetilde C(t)\right)t-3\left(\widetilde E(T)-\widetilde E(t)\right)t^2 \\
c_1&=-\left(\widetilde B(T)-K\widetilde B(t)\right)t^{1/2}+3\left(\widetilde D(T)-\widetilde D(t)\right)t^{3/2} \\
c_2&=-\left(\widetilde C(T)-K\widetilde C(t)\right)t+6\left(\widetilde E(T)-K\widetilde E(t)\right)t^2 \\
c_3&=-\left(\widetilde D(T)-K\widetilde D(t)\right)t^{3/2}\\
c_4&=-\left(\widetilde E(T)-K\widetilde E(t)\right)t^2
\end{aligned}
\end{equation}
We then have the the price of a bond put option is a one--variable third chaos model is
\begin{equation}
p(0,t,T,K)=\frac{1}{\widetilde A(0)\sqrt{2\pi}}\int_{\mathcal{P}^{(4)}_p(z)\geq0}\mathcal{P}^{(4)}_p(z)e^{-\frac{z^2}{2}}dz,
\end{equation}
which can then be used to obtain the price of a caplet according to \eqref{caplet}. As before, once we calculate the roots of the polynomial $\mathcal{P}^{(4)}_p(z)$, both of these expressions reduce to explicit formulas in terms of the standard normal cumulative distribution and density functions (see \cite{Tsujimoto10}).

Regarding the price of a payer swaption, we have that 
\begin{equation}
Z_{tt}-Z_{tT_n}-K\sum_{i=1}^n (T_i-T_{i-1})Z_{tT_i}=\mathcal{P}^{(4)}_{Swp}(z)
\end{equation}
where $z=W_t/\sqrt{t}$ a standard normal random variable and 
\begin{equation}
\mathcal{P}^{(4)}_{Swp}(z):=d_0+d_1 z+d_2 z^2+d_3 z^3+d_4 z^4
\end{equation} 
is a forth degree polynomial with coefficients
\begin{equation}
\begin{aligned}
d_0=&\left(\widetilde A(t)-\widetilde A(T_n)-K\sum_{i=1}^n (T_i-T_{i-1})\widetilde A(T_i)\right)\\
&-\left(\widetilde C(t)-\widetilde C(T_n)-K\sum_{i=1}^n (T_i-T_{i-1})\widetilde C(T_i)\right)t \\
&+3\left(\widetilde E(t)-\widetilde E(T_n)-K\sum_{i=1}^n (T_i-T_{i-1})\widetilde E(T_i)\right)t^2\\
d_1=&\left(\widetilde B(t)-\widetilde B(T_n)-K\sum_{i=1}^n (T_i-T_{i-1})\widetilde B(T_i)\right)t^{1/2} \\
&-3\left(\widetilde D(t)-\widetilde D(T_n)-K\sum_{i=1}^n (T_i-T_{i-1})\widetilde D(T_i)\right)t^{3/2} \\
d_2=&\left(\widetilde C(t)-\widetilde C(T_n)-K\sum_{i=1}^n (T_i-T_{i-1})\widetilde C(T_i)\right)t \\
&-6\left(\widetilde E(t)-\widetilde E(T_n)-K\sum_{i=1}^n (T_i-T_{i-1})\widetilde E(T_i)\right)t^2 \\
d_3=&\left(\widetilde D(t)-\widetilde D(T_n)-K\sum_{i=1}^n (T_i-T_{i-1})\widetilde D(T_i)\right)t^{3/2}\\
d_4=&\left(\widetilde E(t)-\widetilde E(T_n)-K\sum_{i=1}^n (T_i-T_{i-1})\widetilde E(T_i)\right)t^2 
\end{aligned}
\end{equation}
It then follows from \eqref{swaption} that the price of a payer swaption in a one--variable third chaos model is
\begin{equation}
\label{swaption_one_third}
\text{Swp}(0,{\cal T},{\cal N},K)=\frac{1}{\widetilde A(0)\sqrt{2\pi}}\int_{\mathcal{P}^{(4)}_{Swp}(z)\geq0}\mathcal{P}^{(4)}_{Swp}(z)e^{-\frac{z^2}{2}}dz.
\end{equation}
Once more, after we find the roots of $\mathcal{P}^{(4)}_{Swp}(z)$, expression \eqref{swaption_one_third} reduces to an explicit formula in terms of the standard normal cumulative distribution and density functions (see \cite{Tsujimoto10}). 

\subsection{Benchmark models}
\label{benchmark}

In this section we describe the models we use for comparing the calibration performance of chaotic models to option prices. 

We start with the simplest short--rate model for which the initial term structure can be freely specified in order to match observed data, namely the Hull--White 
\cite{HullWhite90} dynamics 
\begin{equation}
dr_t=(\theta(t)-\kappa r_t)dt+\eta dW_t,
\end{equation}
where we choose the function $\theta(t)$ in such a way that the initial term structure for forward rates has the Svensson form in \eqref{descriptive}, so that there are 8 parameters in total to be calibrated for this model. The prices for bond options, caplets and swaptions in a Hull--White model are given by well--known analytic expressions (see for example \cite[pages 76--77]{BrigoMercurio06}). 

Next we consider the rational lognormal model introduced in \cite{FlesakerHughston96} (see related literature in \cite{Goldberg98}, \cite{NakamuraYu00} and \cite{RapisardaSilvotti01}). This is one of the earliest examples of a model in the potential approach and consists of setting 
\begin{equation}
\sigma^2_s=g_1(s)M_s+g_2(s), \quad 0\leq s\leq t,
\end{equation}
where $g_1,g_2$ are nonnegative deterministic functions of time and $M_t$ is a strictly positive continuous martingale such that $M_0=1$. We then see that 
\begin{equation}
Z_{tT}=\int^\infty_T M_{ts}ds=\int^\infty_T E\left[\left.\sigma^2_s\right|{\cal F}_t\right]ds=G_1(T)M_t+G_2(T),
\end{equation}
where
\begin{equation}
G_1(t):=\int^\infty_tg_1(s)ds,\quad\text{and}\quad G_2(t):=\int^\infty_tg_2(s)ds.
\end{equation}
This implies that bond prices  are represented in the following way:
\begin{equation*}
P_{tT}=\frac{G_1(T)M_t+G_2(T)}{G_1(t)M_t+G_2(t)},\quad\text{for}\quad 0\leq t\leq T<\infty.
\end{equation*}
As shown in \cite{FlesakerHughston96}, caplets and swaptions can be priced analytically in a rational lognormal model. Note however that, as also discussed in \cite{FlesakerHughston96} and \cite{Cairns04}, for each fixed $t$ and $T$, bond prices in this model are bounded above and below by the deterministic ratios 
$G_1(T)/G_1(t)$ and $G_2(T)/G_2(t)$. As shown in \cite{Cairns04}, it follows that the short rate $r_t$ is also bounded above and below by the deterministic ratios 
$G^\prime_1(t)/G_1(t)$ and $G^\prime_2(t)/G_2(t)$. As a consequence, rational lognormal models are unable to accurately price deep in the money and out of the money interest rate options.

To implement the rational lognormal model, we follow \cite{NakamuraYu00} and parametrize the functions $g_1$ and $g_2$ as
\begin{equation}
g_1(t)=-k_1\frac{\partial P_{0t}}{\partial t}(P_{0t})^{k_2}
\quad\text{and}\quad
g_2(t)=-\frac{\partial P_{0t}}{\partial t}[1-k_1(P_{0t})^{k_2}],
\quad \text{for}\quad t\geq 0,
\end{equation}
for constants $k_1,k_2\in\mathbb{R}$, where we choose the function  $P_{0t}$ is then chosen so that the initial term structure of forward rates has the Svensson form in \eqref{descriptive}.  Regarding the process $M_t$, we follow \cite{Goldberg98} and represent is as an exponential martingale of the form 
\begin{equation}
M_t=\exp\Big[\eta W_t-\frac{1}{2}\eta^2 t\Big],
\quad \text{for}\quad t\geq 0,
\end{equation}
for some constant $\eta\in\mathbb{R}$. With this choices, there are 9 parameters in total to be calibrated in this model.

The last benchmark that we consider are LIBOR market models. We first introduce the following notation: 
\begin{equation*}
F_j(t):=F_{tT_{j-1}T_j},\quad j=1,2,\dots, 
\end{equation*}
where $\{T_0,T_{1},\dots,T_{i-1},T_i,\dots, T_n\}$ is an increasing set of dates and $F(t,S,T)$ is the forward LIBOR rate defined in \eqref{forward_libor}. 

The lognormal forward market model (LFM), consists of specifying the dynamics:
\begin{equation}
dF_j(t)=\sigma_j(t)F_j(t)dZ^j(t),
\end{equation}
where $\sigma_j(t), t\geq0$ is a deterministic process, and $Z^j$ is a Brownian Motion under the forward measure $\mathbb{Q}^j$, that is, the pricing measure associated with the numeraire $P(\cdot, T_j)$ (see \cite{GemanKarouiRochet95}), with a correlation given by
\begin{equation}
dZ^i(t)dZ^j(t)=\rho_{ij} dt.
\end{equation}
The motivation for introducing such models is that caplets can then be exactly priced using the Black formula introduced in \eqref{black}. Following
\cite[page 224]{BrigoMercurio06}, we parametrize the volatilities as
\begin{equation}
\sigma_i(t)=b_1+(b_2+b_3(T_{i-1}-t))e^{-c_1(T_{i-1}-t)},
\end{equation}
for $b_1, b_2, b_3, c_1\in \mathbb{R}$. Using the Svensson form in \eqref{descriptive} for the initial term structure of forward rates, we see that there are 10 parameters to be calibrated to yield curves and caplet prices alone. 

To price a swaption maturing at $t$ with tenor $(T_n-t)$ we use the Rebonato approximation (see \cite[page 283]{BrigoMercurio06}) for the implied volatility 
\begin{equation}
\label{rebo}
v_{t,T_n}=\sqrt{\frac{1}{T_n}\sum_{i,j=1}^n\frac{w_i(0)w_j(0)F_i(0)F_j(0)}{S(0,t,T_n)^2}\rho_{ij}\int^{T_n}_0\sigma_i(t)\sigma_j(t)dt},
\end{equation}
where the forward swap rates are assumed to be expressed as a linear combination of forward LIBOR rates as follows:
\begin{equation}
S(t,T_0,T_n)=\sum_{i=1}^nw_i(t)F_i(t) \quad \text{where}\quad
w_i(t)=\frac{(T_i-T_{i-1})P_{tT_i}}{\sum_{k=1}^n(T_k-T_{k-1})P_{tT_k}}.
\end{equation}
For the correlations appearing in \eqref{rebo}, we use the Schoenmakers and Coffey \cite{SchoenmakersCoffey06} parametrization:
\begin{equation}
\begin{split}
\rho_{ij}=
\exp\Big[&-\frac{|j-i|}{n-1}\Big(-\log\rho_\infty+\eta_1\frac{i^2+j^2+ij-3ni-3nj+3i+3j+2n^2-n-4}{(n-2)(n-3)}\\
&-\eta_2\frac{i^2+j^2+ij-ni-nj-3i-3j+3n+2}{(n-2)(n-3)}\Big)\Big],
\end{split}
\end{equation}
for $i,j=1,2,\dots,n$ and constants $\eta_1,\eta_2,\rho_\infty\in\mathbb{R}$ satisfying the conditions $\eta_2\geq 0$ and $0\leq\eta_1+\eta_2\leq-\log\rho_\infty$. As we can see, if we use either only yields and caplets, we need to calibrate 10 parameters, whereas there are in total 13 parameters to be calibrated to yields and swaptions or yields, caplets, and swaptions together.

\subsection{Calibration Results}

We describe the details of the calibration to yields and option prices in Appendix \ref{option_procedure}. We calibrate 6 different chaos models using two distinct data set from the UK interest rate market: weekly observations of yields and implied volatilities for ATM caplets and ATM swaptions from September 2000 to August 2001 and from May 2005 to May 2006. For comparison, we also calibrate the three benchmark models described in the Section \ref{benchmark}: the Hull and White short rate model, the rational lognormal model and the lognormal forward LIBOR market model. 

\begin{figure}[h]

\begin{center}
\includegraphics[width=\textwidth]{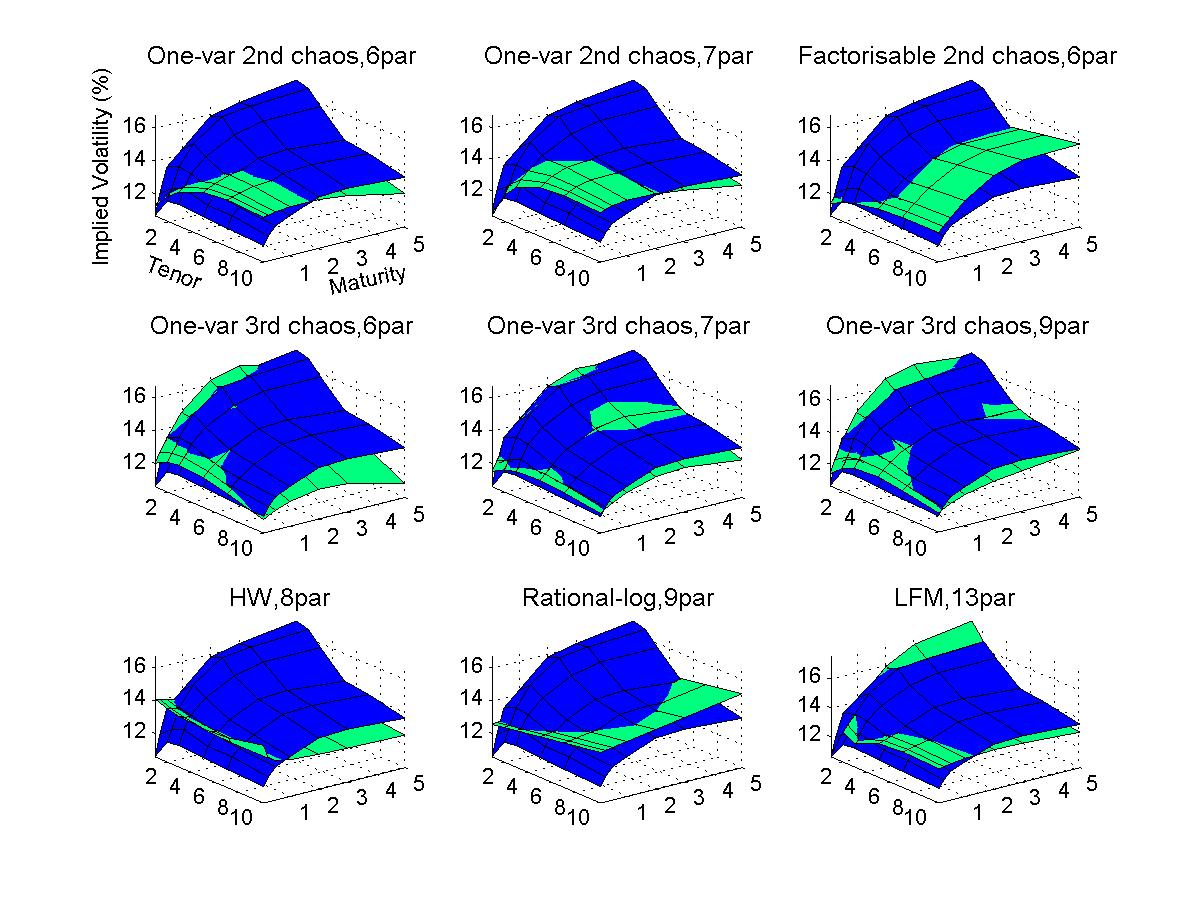}
\end{center}
\caption{Implied volatility for ATM swaptions  on October 20th, 2000  (Blue: Market Quotes, Green: Theoretical Values).}
\label{figureD1}

\begin{center}
\includegraphics[width=\textwidth]{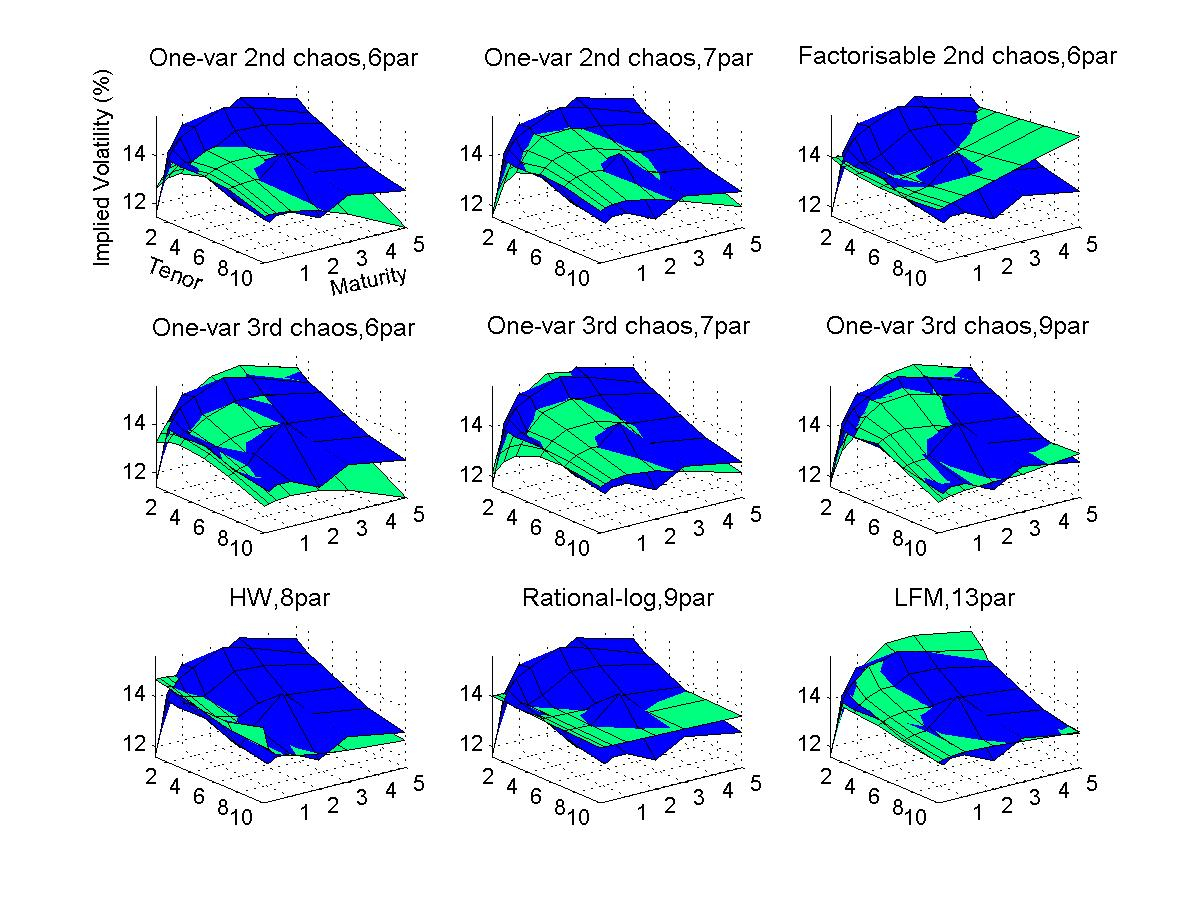}
\end{center}
\caption{Implied volatility for ATM swaptions  on August 12th, 2005  (Blue: Market Quotes, Green: Theoretical Values).}
\label{figureD2}
\end{figure}

\begin{figure}[h]
\begin{center}
\includegraphics[width=\textwidth]{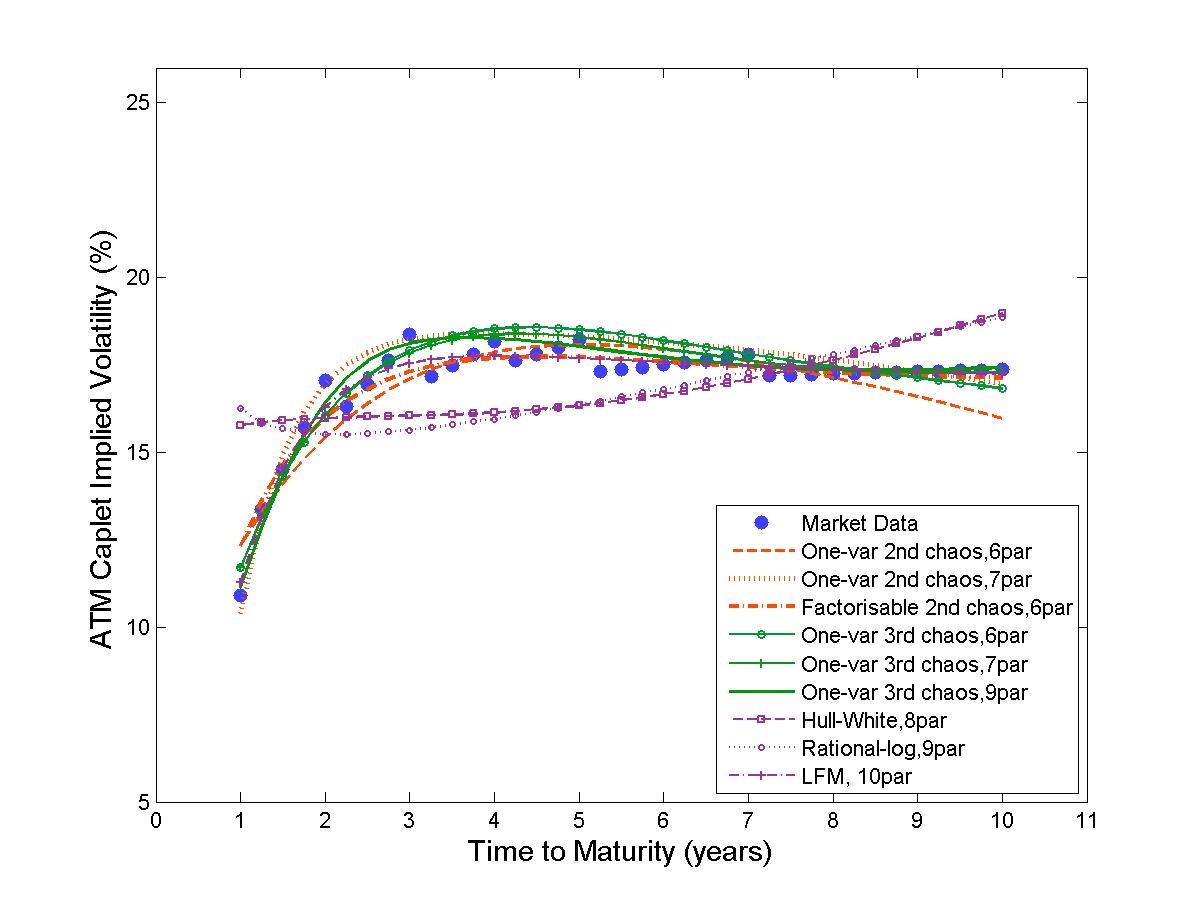}
\end{center}
\caption{Implied volatility for ATM caplets on December 8th, 2000.}
\label{figureD3}

\begin{center}
\includegraphics[width=\textwidth]{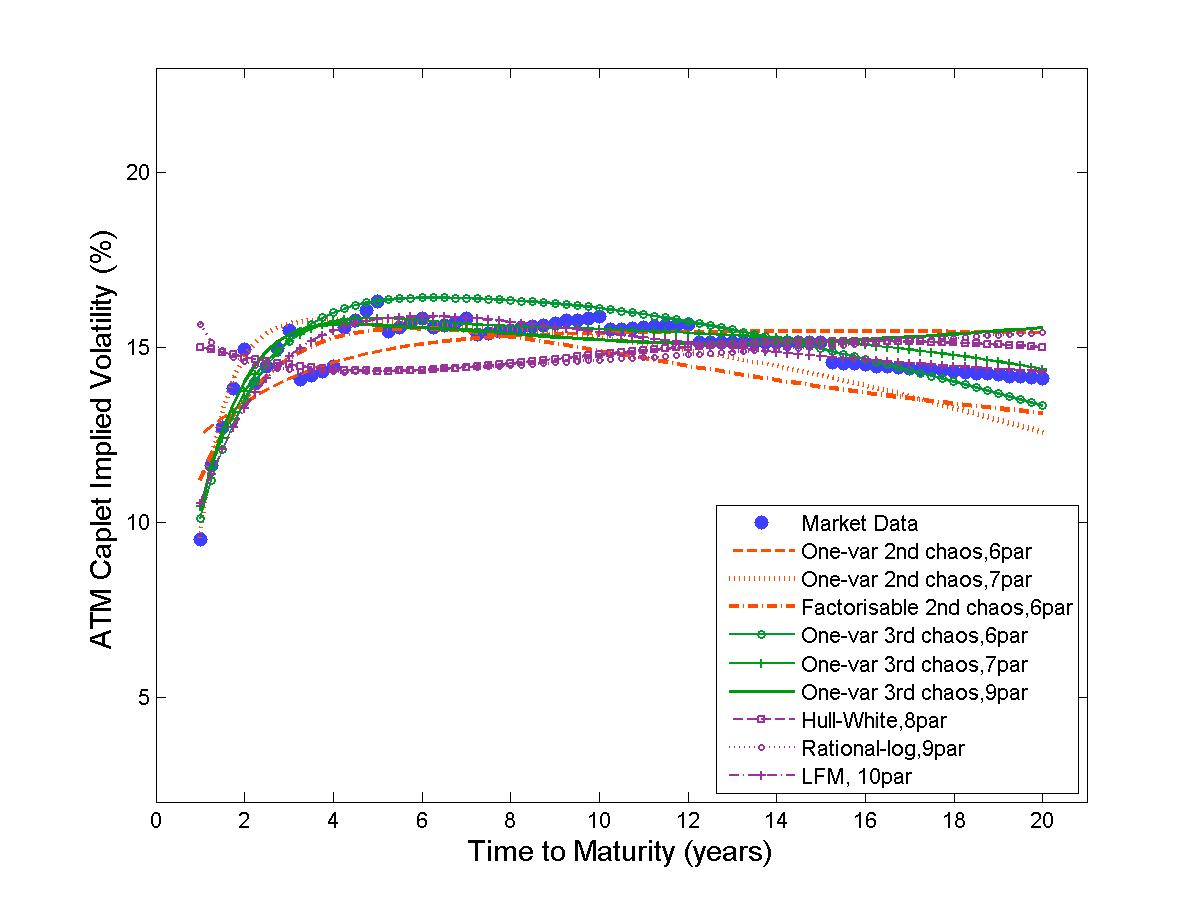}
\end{center}
\caption{Implied volatility for ATM caplets on February 10th, 2006.}
\label{figureD4}
\end{figure}

We can gain a first impression in Figures \ref{figureD1} and \ref{figureD2}, which show the calibrated implied volatility surfaces for ATM swaptions of different maturities and tenors at two specific calibration dates, one from each of the data sets above. We can see that neither the Hull and White nor the rational lognormal models have enough flexibility to reproduce the market quotes, with the LIBOR and chaos models providing much better fits. Figures \ref{figureD3} and 
\ref{figureD4} show similar results for the term structure of ATM caplets. In particular, we observe that both the Hull and White and the rational lognormal fail to capture the hump shaped term structure of caplet implied volatility, whereas the LIBOR and chaos models are able to do so.

\begin{figure}[h]

\begin{center}
\includegraphics[width=\textwidth]{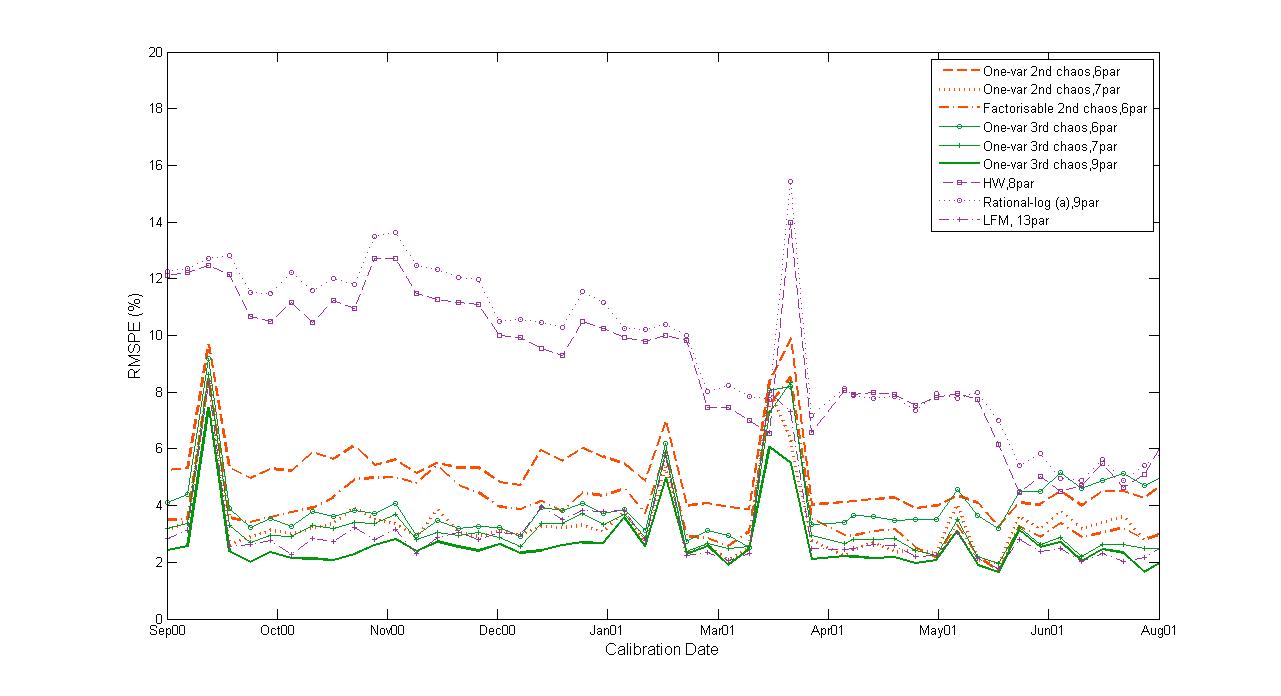}
\end{center}
\caption{Total error for yields and ATM caplets in 2000--2001.}
\label{figure1}

\begin{center}
\includegraphics[width=\textwidth]{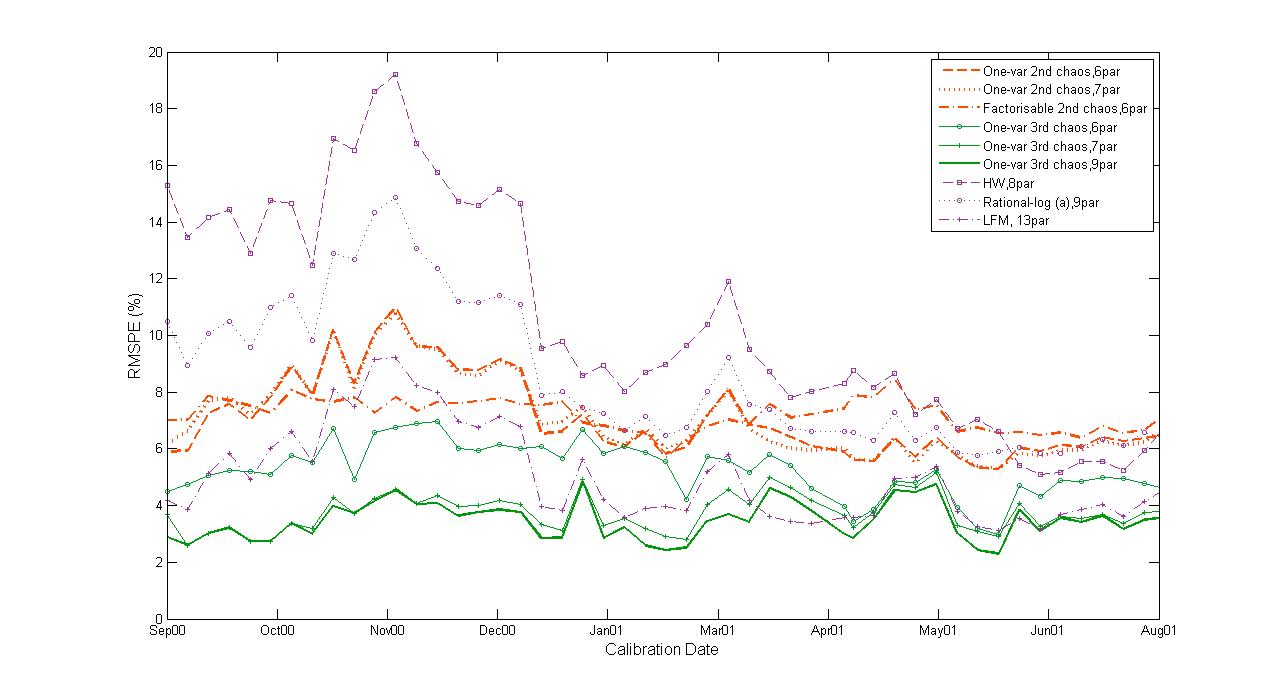}
\end{center}
\caption{Total error for yields and ATM swaptions in 2000--2001.}
\label{figure2}

\begin{center}
\includegraphics[width=\textwidth]{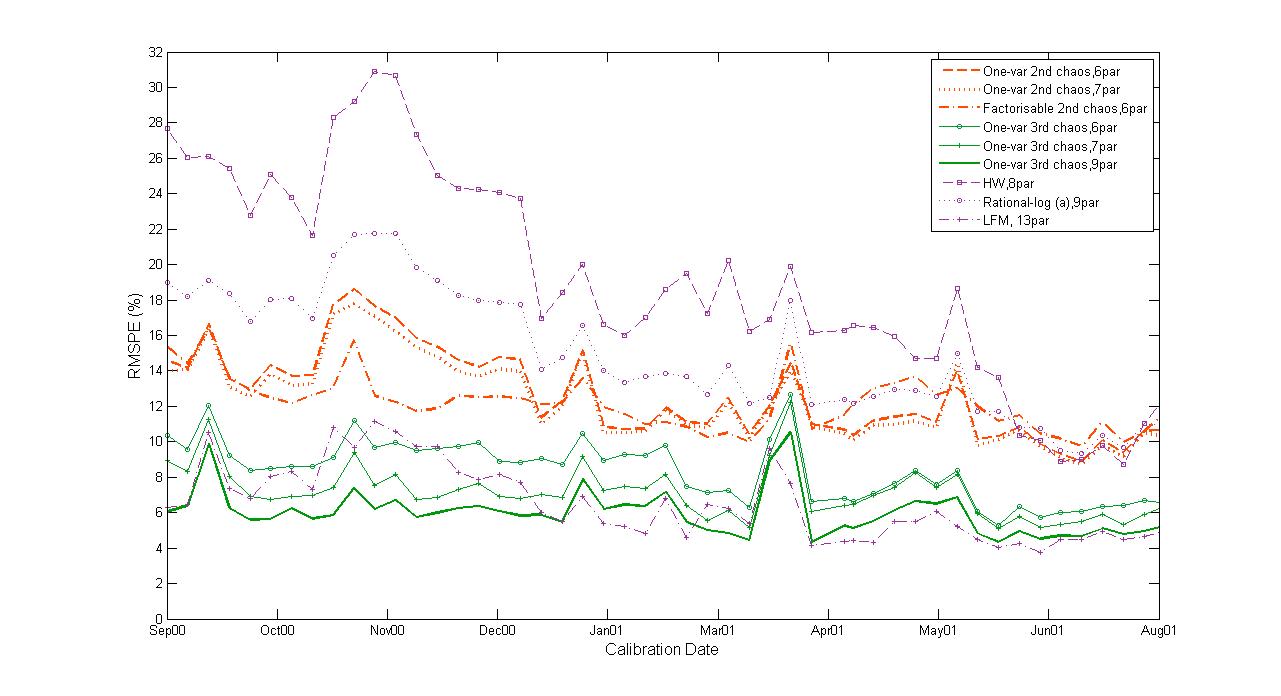}
\end{center}
\caption{Total error for yields, ATM caplets and ATM swaptions in 2000--2001.}
\label{figure3}

\end{figure}

\begin{figure}[h]

\begin{center}
\includegraphics[width=\textwidth]{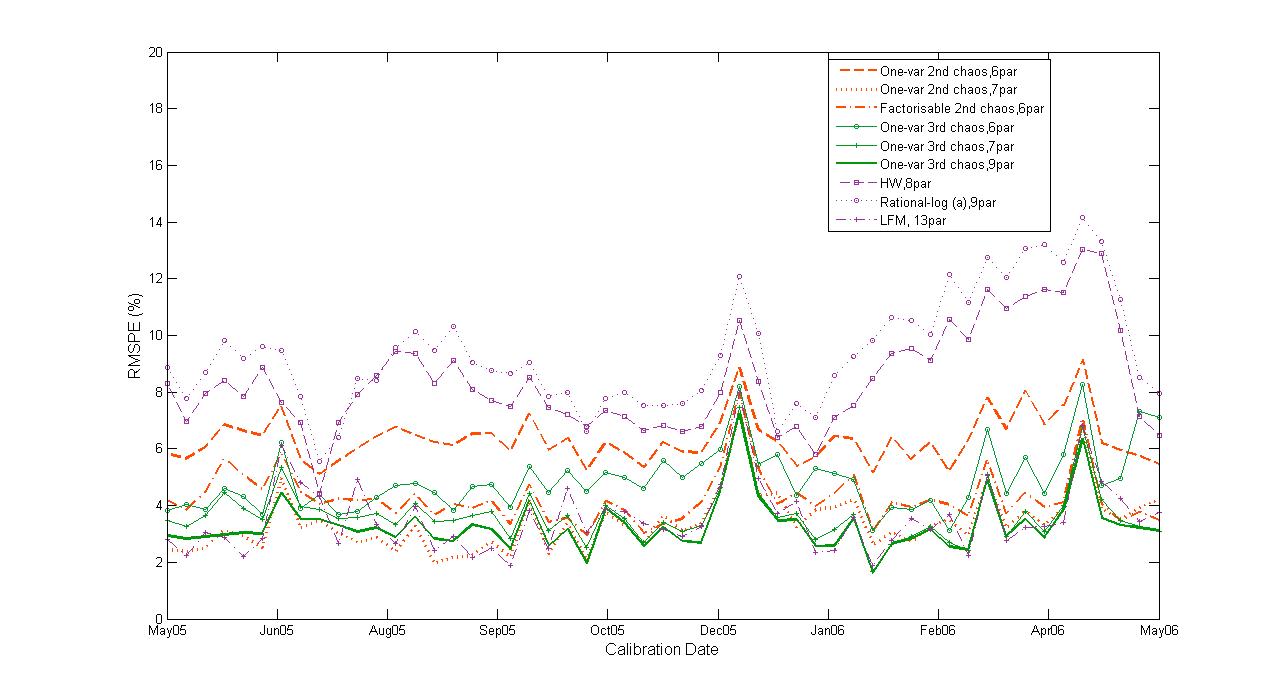}
\end{center}
\caption{Total error for yields and ATM caplets in 2005--2006.}
\label{figure4}

\begin{center}
\includegraphics[width=\textwidth]{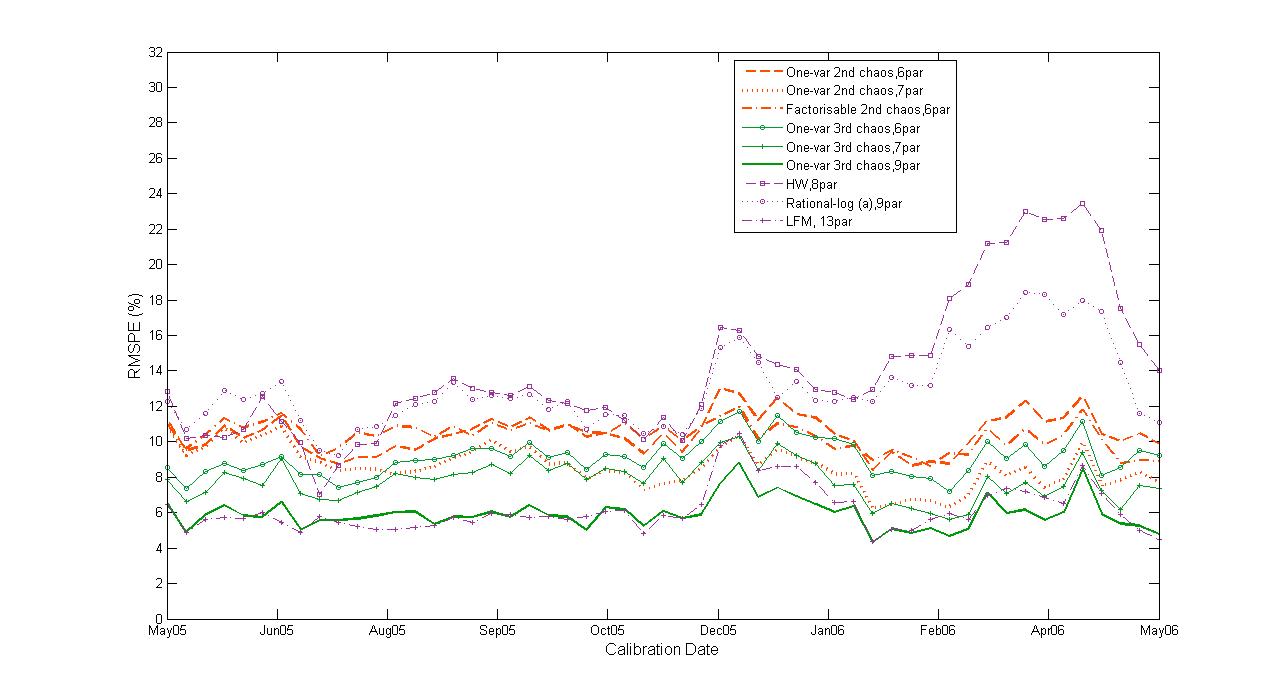}
\end{center}
\caption{Total error for yields and ATM swaptions in 2005--2006.}
\label{figure5}

\begin{center}
\includegraphics[width=\textwidth]{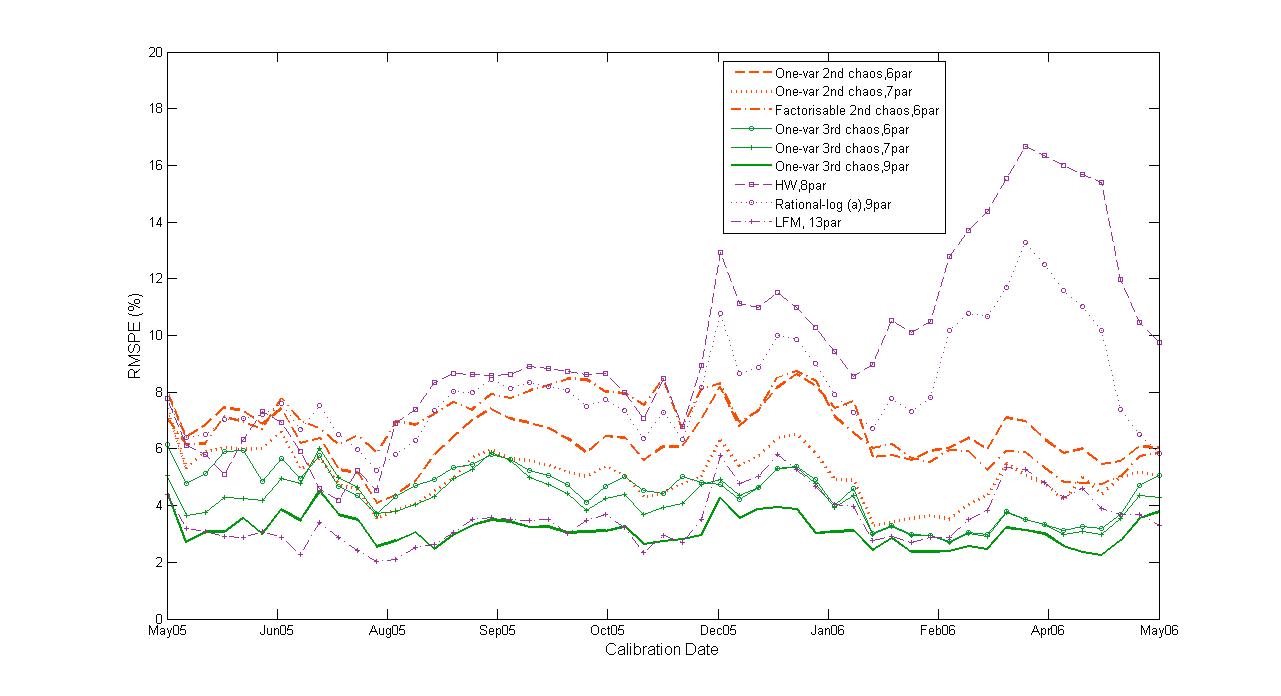}
\end{center}
\caption{Total error for yields, ATM caplets and ATM swaptions in 2005--2006.}
\label{figure6}

\end{figure}

The quantitative calibration results are summarized results in Tables \ref{table:3} to \ref{table:8}.  The first column in each table labels the calibrated chaos models as listed in Appendix \ref{option_procedure} and the benchmark models as listed in Section \ref{benchmark}. The second column characterizes the type of model, whereas the third one gives the number of calibrated parameters. The remaining columns in each table show the errors defined in Appendix \ref{option_procedure} averaged across the different calibration dates. Specifically, the fourth column in each table shows the average value for the objective function minimized in the calibration. We plot this error across time for all calibration dates in Figures \ref{figure1} to \ref{figure6}. The next columns show the average contribution to this total error coming from yields and the options used in each calibration. In the last column of Tables  \ref{table:3}, \ref{table:4}, \ref{table:6} and \ref{table:7}, we also show the correspond to the pricing error obtained when we use the calibrated parameters to forecast the prices of options that were not used in the calibration. For example, in Table \ref{table:3} we calibrate the model parameters to yields and caplets and use them to forecast the prices of swaptions.

We can readily see that in general all chaos models that we consider fit the observed data better than the Hull and White and the rational lognormal models, as well as having smaller pricing errors for the options that were not used in each particular calibration. When compared with the LIBOR model, however, we see that only third chaos models have competitive performance. 

For a more detailed comparison, we consider relative frequencies for model selection based on the Akaike Information Criterion described in Appendix \ref{option_procedure}. For each of our two data sets, we compared the relative performance of third chaos models with 7 and 9 parameters and the benchmark LIBOR model. The results are presented in Tables \ref{AIC1} and \ref{AIC2}. Whereas the relative performance of the third chaos model with 7 parameters is somewhat ambiguous and depends on which type of options are considered, we find that the third chaos model with 9 parameters consistently outperforms the lognormal forward LIBOR market model with 13 parameters. For example, in the first data set, it outperforms LIBOR in 36 out of the 53 calibration dates when we use yields and caplets, 53 out of 53 calibration dates when we use yields and swaptions, and 39 out of 53 calibration dates when we use yields, caplets and swaptions. For the second data set, the corresponding number of dates are 52, 44 and 39 respectively.

\newpage
\begin{table}[h]
\caption{Yield and ATM caplet calibration for 2000-2001}
\centering
\begin{tabular}{|c|l|r|r|r|r|r|}
\hline
No. & Model & N & TotalE1 (\%)& YldE (\%)& CplE (\%)&  SwpE (\%)  \\
\hline
1 & one-var 2nd chaos & 6 & 5.1 & 2.0 & 4.6  & 14.9 \\
\hline
2 & one-var 2nd chaos & 7 & 3.3 & 1.7 & 2.7  & 16.3 \\
\hline
3 & factorizable 2nd & 6 & 3.8 & 2.1 & 3.1  & 26.5 \\
\hline
4 & one-var 3rd chaos & 6 & 4.2 & 2.0 & 3.5  & 15.5 \\
\hline
5 & one-var 3rd chaos & 7 & 3.2 & 1.3 & 2.9  & 15.7 \\
\hline
6 & one-var 3rd chaos & 9 & 2.6 & 1.1 & 2.3  & 17.0 \\
\hline
I & Hull-White & 8 & 8.7 & 0.6 & 8.7  & 25.8 \\
\hline
II & Rational-log  & 9 & 9.2 & 0.6 & 9.2  & 13.9 \\
\hline
III & LIBOR & 10 & 3.0 & 0.6 & 3.0  & -  \\
\hline
\end{tabular}  
\label{table:3}

\caption{Yield and ATM swaption calibration for $2000-2001$}
\centering
\begin{tabular}{|c|l|r|r|r|r|r|}
\hline
No. & Model & N & TotalE2 (\%)& YldE (\%)& CplE (\%)&  SwpE (\%)  \\
\hline
1 & one-var 2nd chaos & 6 & 7.1 & 1.8 & 6.8  & 14.5 \\
\hline
2 & one-var 2nd chaos & 7 & 7.1 & 2.0 & 6.7  & 14.6 \\
\hline
3 & factorizable 2nd & 6 & 7.1 & 2.1 & 6.8  & 14.3 \\
\hline
4 & one-var 3rd chaos & 6 & 5.3 & 2.9 & 4.1  & 10.2 \\
\hline
5 & one-var 3rd chaos & 7 & 3.8 & 1.5 & 3.4  & 8.6 \\
\hline
6 & one-var 3rd chaos & 9 & 3.5 & 1.5 & 3.1  & 9.1 \\
\hline
I & Hull-White & 8 & 10.2 & 0.6 & 10.2  & 17.6 \\
\hline
II & Rational-log & 9 & 8.4 & 0.6 & 8.4  & 15.3 \\
\hline
III & LIBOR & 13 & 5.0 & 0.6 & 5.0  & 8.1 \\
\hline
\end{tabular}  
\label{table:4}

\caption{Yield, ATM caplet and ATM swaption calibration for $2000-2001$}
\centering
\begin{tabular}{|c|l|r|r|r|r|r|}
\hline
No. & Model & N & TotalE2 (\%)& YldE (\%)& CplE (\%)&  SwpE (\%)  \\
\hline
1 & one-var 2nd chaos & 6 & 12.5 & 2.2 & 9.3  & 7.9 \\
\hline
2 & one-var 2nd chaos & 7 & 12.1 & 2.4 & 9.3  & 7.3 \\
\hline
3 & factorizable 2nd & 6 & 12.1 & 2.6 & 8.4  & 8.2 \\
\hline
4 & one-var 3rd chaos & 6 & 8.2 & 4.3 & 4.4  & 5.2 \\
\hline
5 & one-var 3rd chaos & 7 & 7.1 & 1.6 & 4.4  & 5.2 \\
\hline
6 & one-var 3rd chaos & 9 & 5.9 & 2.2 & 4.1  & 3.4 \\
\hline
I & Hull-White & 8 & 18.4 & 0.6 & 12.2  & 13.7 \\
\hline
II & Rational-log  & 9 & 14.6 & 0.6 & 10.0  & 10.6 \\
\hline
III & LIBOR & 13 & 6.5 & 0.6 & 5.5  & 3.1 \\
\hline
\end{tabular}
\label{table:5}
\end{table}

\newpage
\begin{table}[h]
\caption{Yield and ATM caplet calibration for $2005-2006$}
\centering
\begin{tabular}{|c|l|r|r|r|r|r|}
\hline
No. & Model & N & TotalE3 (\%)& YldE (\%)& CplE (\%)&  SwpE (\%)  \\
\hline
1 & one-var 2nd chaos & 6 & 6.3 & 1.6 & 6.1  & 9.4 \\
\hline
2 & one-var 2nd chaos & 7 & 3.4 & 1.5 & 3.0  & 14.0 \\
\hline
3 & factorizable 2nd & 6 & 4.3 & 2.4 & 3.4  & 20.0 \\
\hline
4 & one-var 3rd chaos & 6 & 4.9 & 1.9 & 4.4  & 26.2 \\
\hline
5 & one-var 3rd chaos & 7 & 3.6 & 1.4 & 3.2  & 14.2 \\
\hline
6 & one-var 3rd chaos & 9 & 3.3 & 1.3 & 3.0  & 17.7 \\
\hline
I & Hull-White & 8 & 8.4 & 0.4 & 8.4  & 16.3 \\
\hline
II & Rational-log  & 9 & 9.4 & 0.4 & 9.4  & 10.5 \\
\hline
III & LIBOR & 10 & 3.5 & 0.4 & 3.5  & - \\
\hline
\end{tabular}
\label{table:6}

\caption{Yield and ATM swaption calibration for $2005-2006$}
\centering
\begin{tabular}{|c|l|r|r|r|r|r|r|r|}
\hline
No. & Model & N & TotalE1 (\%)& YldE (\%)& CplE (\%)&  SwpE (\%)  \\
\hline
1 & one-var 2nd chaos & 6 & 6.5 & 3.2 & 5.5  & 32.2 \\
\hline
2 & one-var 2nd chaos & 7 & 5.0 & 1.5 & 4.8  & 11.9 \\
\hline
3 & factorizable 2nd & 6 & 6.8 & 2.3 & 6.4  & 13.7 \\
\hline
4 & one-var 3rd chaos & 6 & 4.5 & 2.2 & 3.8  & 21.2 \\
\hline
5 & one-var 3rd chaos & 7 & 4.2 & 1.6 & 3.8  & 13.4 \\
\hline
6 & one-var 3rd chaos & 9 & 3.1 & 1.3 & 2.8  & 14.2 \\
\hline
I & Hull-White & 8 & 9.5 & 0.4 & 9.5  & 11.2 \\
\hline
II  & Rational-log  & 9 & 8.2 & 0.4 & 8.2  & 10.9 \\
\hline
III & LIBOR & 13 & 3.5 & 0.4 & 3.5  & 14.8 \\
\hline
\end{tabular}  
\label{table:7}

\caption{Yield, ATM caplet and ATM swaption calibration for $2005-2006$}
\centering
\begin{tabular}{|c|l|r|r|r|r|r|r|r|}
\hline
No. & Model & N & TotalE1 (\%)& YldE (\%)& CplE (\%)&  SwpE (\%)  \\
\hline
1 & one-var 2nd chaos & 6 & 10.4 & 2.5 & 7.3  & 6.9 \\
\hline
2 & one-var 2nd chaos & 7 & 8.6 & 1.5 & 6.3  & 5.6 \\
\hline
3 & factorizable 2nd & 6 & 10.3 & 1.9 & 7.9  & 6.2 \\
\hline
4 & one-var 3rd chaos & 6 & 9.1 & 3.3 & 5.8  & 6.1 \\
\hline
5 & one-var 3rd chaos & 7 & 7.8 & 1.8 & 5.0  & 5.5 \\
\hline
6 & one-var 3rd chaos & 9 & 5.9 & 1.4 & 4.1  & 4.0 \\
\hline
I & Hull-White & 8 & 14.0 & 0.4 & 10.1  & 9.5 \\
\hline
II & Rational-log  & 9 & 13.0 & 0.4 & 8.4  & 9.9 \\
\hline
III & LIBOR & 13 & 6.2 & 0.4 & 4.8  & 3.8 \\
\hline
\end{tabular}
\label{table:8}
\end{table}

\begin{table}[h]
\caption{AIC model selection relative frequency (first dataset)}
\centering
\vspace{0.3cm}
\begin{tabular}{|c|c|c|c|c|}
\hline
&Model & Caplets &  Swaptions & Joint \\
\hline
\multirow{2}{*}{comparison 1}&one-var $3$rd chaos ($7$ par) & 2/53 & 50/53  & 23/53 \\
&LIBOR (13 par)& 51/53 & 3/53  & 30/53 \\
\hline
\multirow{2}{*}{comparison 2}& one-var $3$rd chaos ($9$ par) & 36/53  & 53/53  & 39/53 \\
& LIBOR (13 par) & 17/53  & 0/53  & 14/53 \\
\hline
\end{tabular}  
\label{AIC1}
\end{table}

\begin{table}[h]
\caption{AIC model selection relative frequency (second dataset)}
\centering
\vspace{0.3cm}
\begin{tabular}{|c|c|c|c|c|}
\hline
&Model & Caplets &  Swaptions & Joint \\
\hline
\multirow{2}{*}{comparison 1}&one-var $3$rd chaos ($7$ par) & 14/53 & 23/53  & 7/53 \\
&LIBOR & 39/53 & 30/53  & 46/53 \\
\hline
\multirow{2}{*}{comparison 2}& one-var $3$rd chaos ($9$ par) & 52/53  & 44/53  & 39/53 \\
& LIBOR & 1/53  & 9/53  & 14/53 \\
\hline
\end{tabular}  
\label{AIC2}
\end{table}

\section{Concluding remarks}
\label{conclusion}

We proposed and implemented a systematic way to calibrate chaotic models for interest rates to available term structure and option data. The calibration performance to initial term structures is comparable to that of traditional descriptive models for forward rates with the same number of parameters, with the advantages of guaranteed positivity and consistency with a fully stochastic model for the time evolution of interest rates. When we include option data in the form of at-the-money caplets and swaptions, we see that chaos models perform significantly better than the benchmark Hull and White and rational lognormal models, and comparably to a LIBOR market model with a higher number of parameters. If we take the number of parameters into account, a conservative information criterion shows that one of our chaos models consistently outperforms the benchmark LIBOR market model.

The underlying reason for the superior performance of LIBOR market models when compared, for example, to Markovian short--rate models is the rich correlation structure they provide for LIBOR forward rates at different dates. Similarly, in chaos models the resulting short rate is in general not Markovian, and our calibration results show that an equally rich correlation structure can be achieved without having to model forward rates individually under each corresponding forward measure. 

The next step in our program is to consider options that are not at-the-money. The current industry standard for this is to use stochastic volatility models for either forward or swap rates. In a paper in preparation we show that chaos models naturally give rise to rates with stochastic volatility, and explore this fact to calibrate them to the smile observe in interest rate data.

Expectedly, much work remains to be done, in particular on the interpretation of the financial interpretation of the chaos coefficients, where techniques such as principal components analysis might shed some additional light on the meaning of the most relevant calibrated parameters. Extensions to other classes of financial products, notably exchange rate derivatives, are also possible. We believe the analysis presented in this paper demonstrates the viability of chaos models as serious contenders for practical use in the financial industry and will stimulate further work in the area.

\section*{Acknowledgements} We are grateful for helpful comments by M. Avellaneda, D. Brody, D. Crisan, J-P. Fouque, L. Hughston, T. Hurd, J. Teichmann, M. Yor and the participants at the CRFMS seminar, Santa Barbara, May 2009, the Mathematical Finance and Related Topics in Engineering and Economics Workshop, Kyoto, August 2009, the Research in Options Conference, Buzios, November 2009, the Sixth World Congress of the Bachelier Finance Society, Toronto, June 2010, and the Imperial College finance seminar, October 2010, where this work was presented.

\appendix
\section{Term structure calibration procedure}
\label{term_procedure}

We begin by listing our choices of chaos coefficients of different orders according to the general parametric form in \eqref{parametric}. We start with first chaos models directly inspired by the Nelson--Siegel and Svensson parametric forms in \eqref{descriptive}, namely:
\begin{align}
\phi_1(s)&= (b_1+b_2s)e^{-c_1s} \label{A1}\\
\phi_1(s)&= (b_1+b_2s)e^{-c_1s}+b_3se^{-c_2s}. \\
\intertext{Moving up one chaos order, we consider next one--variable second chaos models with}
\alpha(s)&=(b_1+b_2s)e^{-c_1s}, \quad \beta(s)=(b_3+b_4s)e^{-c_2s} \\
\alpha(s)&=b_1e^{-c_1s}, \quad \beta(s)=(b_2+b_3s)e^{-c_2s}+b_4se^{-c_3s}\\
\alpha(s)&=(b_1+b_2s)e^{-c_1s}+b_3se^{-c_2s}, \quad \beta(s)= b_4se^{-c_3s}\\
\intertext{as well as factorizable second chaos models with} 
\alpha(s)&=b_1e^{-c_1s}, \quad \beta(s)=b_2e^{-c_2s},\quad \gamma(s)=(1+b_3)e^{-c_3s} \\
\alpha(s)&=b_1e^{-c_1s}, \quad \beta(s)=(b_2+b_3s)e^{-c_2s},\quad \gamma(s)=e^{-c_3s} \\
\alpha(s)&=(b_1+b_2s)e^{-c_1s}, \quad \beta(s)=b_3e^{-c_2s},\quad \gamma(s)=e^{-c_3s} \\
\alpha(s)&=b_1e^{-c_1s}, \quad \beta(s)=(b_2+b_3s)e^{-c_2s},\quad \gamma(s)=(1+b_4s)e^{-c_3s} \\
\alpha(s)&=(b_1+b_2s)e^{-c_1s}, \quad \beta(s)=b_3e^{-c_2s},\quad \gamma(s)=(1+b_4s)e^{-c_3s}.\\ 
\intertext{Finally, we consider the following one--variable third chaos models:}
\alpha(s)&=b_1e^{-c_1s}, \quad \beta(s)=b_2e^{-c_2s},\quad \delta(s)=b_3e^{-c_3s} \\
\alpha(s)&=b_1e^{-c_1s}, \quad \beta(s)=b_2e^{-c_2s},\quad \delta(s)=(b_3+b_4s)e^{-c_3s} \\
\alpha(s)&=b_1e^{-c_1s}, \quad \beta(s)=(b_2+b_3s)e^{-c_2s},\quad \delta(s)=b_4e^{-c_3s} \\
\alpha(s)&=(b_1+b_2s)e^{-c_1s}, \quad \beta(s)=b_3e^{-c_2s},\quad \delta(s)=b_4e^{-c_3s} \label{A14} 
\end{align}

We now describe in detail the steps we take to calibrate these chaos models to observed yield curves, which we obtain from clean prices of treasury coupon strips in the UK bond market from the United Kingdom Debt Management Office (DMO) \footnote{http://www.dmo.gov.uk/index.aspx?page=Gilts/Daily} according to
\begin{equation}
y_{0T}:=-\frac{1}{T}\log P_{0T},
\end{equation} 
using an Actual/Actual day-count convention \cite{BrigoMercurio06}. We consider the following two data sets:
\begin{enumerate}
\item  Bond prices at $146$ dates (every other business day) from January $1998$ to January $1999$, with around 49 to 62 maturities for each date. 
\item  Bond prices at $157$ dates (every Friday) from December $2002$ to December $2005$, with around 100 to 130 maturities for each date. 
\end{enumerate}
Note that the first dataset contains a volatile market including the period of the Long-Term Capital Management (LTCM) crisis, whereas the second dataset corresponds to more moderate market conditions.

We then apply the maximum likelihood estimation (MLE) method suggested by Cairns in \cite{Cairns98} \cite{CairnsPritchard01} to each of the models 
\eqref{A1}--\eqref{A14} and each starting date in the data sets above. This is done as follows: given a model with parameter vector 
$\Theta=(b_1,\ldots,b_{N_1},c_1,\ldots,c_{N_2})$ and a starting date $t_0=0$ with available maturities $T_i$, $i=1,\ldots,n$, we denote the theoretical prices by $P_{0T_i}(\Theta)$ and the corresponding observed bond prices by $\overline{P}_{0T_i}$. We then assume that
\begin{equation}
\log\overline{P}_{0T_i}\sim \mathcal{N}(\log P_{0T_i}(\Theta),\nu^2(P_{0T_i}(\Theta),d_i)),
\end{equation}
where $d_i$ is the Macaulay duration for the bond and
\begin{equation}
\nu^2(p,d)=\frac{\sigma^2_0(p)[\sigma^2_{\infty}d^2b(p)+1]}{\sigma^2_0(p)d^2b(p)+1},
\quad b(p)=\frac{\sigma_d^2}{\sigma_0^2(p)[\sigma_{\infty}-\sigma_0^2(p)]},
\end{equation}
with the error parameters 
\begin{equation}
\sigma_0(p)=\frac{1}{3200p},\quad \sigma_d=0.0005,\quad \sigma_\infty=0.001,
\end{equation}
adopted in \cite{Cairns98} for the UK bond market data between January $1992$ and November $1996$. As explained in \cite{Cairns98}, $\sigma_0(p)$ is based on the assumption that the published bond prices have rounding error of around $1/32$ per $100$ nominal price, whereas $\sigma_d$ corresponds to the assumption that the difference between actual and expected yields have independent errors of the order of five basis points. Finally, $\sigma_\infty$ places a limit on the magnitude of price errors for long dated bonds. 

This leads to the log-likelihood function
\begin{equation}
\label{loglike}
L(\Theta)=-\frac{1}{2}\sum_{i=1}^{n}\Big[\log[2\pi\nu^2(P_{0T_i}(\Theta),d_i)]+\frac{(\log P_{0T_i}(\Theta)-\log\overline{P}_{0T_i})^2}{\nu^2(P_{0T_i}(\Theta),d_i)}\Big].
\end{equation}
We then use a global search procedure to find the global maximum for the log--likelihood function 
\eqref{loglike}, that is, to avoid finding a local maximum, we repeat the procedure using 1000 different random starting points and select the best maximization result. 

Having estimated the parameter vector $\widehat \Theta$, we denote by $y_{0T_i}(\widehat\Theta)$ the fitted yield for maturity $T_i$ and by $\overline{y}_{0T_i}$ the corresponding observed yield. We then define the fitting Root-Mean-Squared Percentage Error (RMSPE) as
\begin{equation}
\text{RMSPE}=\sqrt{\frac{1}{n}\sum^{n}_{i=1}\Big[\frac{y_{0T_i}(\widehat \Theta)-\overline{y}_{0T_i}}{\bar{y}_{0T_i}}\Big]^2}.
\end{equation}

We then apply the Diebold-Mariano (DM) statistics \cite{DieboldMariano95} based on RMSPE to compare fitting performances as is done in \cite{JarrowLiZhao07} and \cite{TrolleSchwartz09}. Here for the computation we use the program DMARIANO\footnote{http://ideas.repec.org/c/boc/bocode/s433001.html} in the statistics package STATA, with a lag order of thirteen in both of our two data sets. The null hypothesis, which is that two models have the same fitting errors, may be rejected at $5 \%$ level if the absolute value of the DM statistics is greater than $1.96$. We compare the calibration performance of the Chaos Models with a descriptive model for forward rates in Svensson form: the higher the DM statistics, the more a chaotic model outperforms the Svensson model. 

\section{Option calibration procedure}
\label{option_procedure}

For option calibration, we consider one--variable second chaos models with
\begin{align}
\alpha(s)&=(b_1+b_2s)e^{-c_1s}, \quad \beta(s)=(b_3+b_4s)e^{-c_2s} \\
\alpha(s)&=(b_1+b_2s)e^{-c_1s}+b_3se^{-c_2s}, \quad \beta(s)= b_4se^{-c_3s},\\
\intertext{factorizable second chaos models with} 
\alpha(s)&=(b_1+b_2s)e^{-c_1s}, \quad \beta(s)=b_3e^{-c_2s},\quad \gamma(s)=e^{-c_3s} \\
\intertext{and one--variable third chaos models with}
\alpha(s)&=b_1e^{-c_1s}, \quad \beta(s)=b_2e^{-c_2s},\quad \delta(s)=b_3e^{-c_3s} \\
\alpha(s)&=(b_1+b_2s)e^{-c_1s}, \quad \beta(s)=b_3e^{-c_2s},\quad \delta(s)=b_4e^{-c_3s}  \\
\alpha(s)&=(b_1+b_2s)e^{-c_1s}, \quad \beta(s)=(b_3+b_4s)e^{-c_2s},\quad \delta(s)=(b_5+b_6s)e^{-c_3s},
\end{align}
as well as each of the three benchmark models described in Section \ref{benchmark}.

Regarding the data, zero-coupon yields are bootstrapped from the LIBOR, Future and Swap rates (see \cite{Fabozzi02} for the detail of the bootstrapping technique) and interest rate option prices from ICAP (Garban Intercapital - London) and TTKL (Tullett \& Tokyo Liberty - London) via the Bloomberg Database.\footnote{Here we would like to acknowledge helpfulness of the Bloomberg help desk staff, who have aided greatly our understanding of the actual market data. We particularly wish to extend thanks David Culshaw, from ICAP, for his assistance.} We consider the following two data sets from the UK interest rate market:
\begin{itemize}
\item  Data between September $2000$ and August $2001$ at $53$ dates (every Friday closing mid price) consisting of
\begin{itemize}
\item zero-coupon yields with 17 maturities ranging from one month to 20 years,
\item implied volatilities for ATM caplets with 37 maturities ranging from one to 10 years,
\item implied volatilities for ATM swaptions with 7 maturities ranging from one month to 5 years and 6 tenors ranging from one to 10 years.
\end{itemize}
\item  Data between May $2005$ and May $2006$ at $53$ dates (every Friday closing mid price) consisting of 
\begin{itemize}
\item zero-coupon yields with 22 maturities ranging from one month to 20 years.
\item implied volatilities for ATM caplets with 77 maturities ranging from one to 20 years,
\item implied volatilities for ATM swaptions with 7 maturities ranging from one month to 5 years and 6 tenors ranging from one to 10 years.
\end{itemize}
\end{itemize}
Note here that the option data corresponds to a part of the data in \cite{TrolleSchwartz09}, where data was analyzed between August $1998$ and January $2007$. 
We obtain caplet implied volatilities by bootstrapping ATM cap implied volatilities observed in the market using the technique described in \cite{GatarekBachertMaksymiuk06}, where the ATM caplets implied volatilities maturing at six months and nine months are obtained by constant extrapolation. The extrapolation is necessary to bootstrap the other ATM caplet implied volatilities, but when we calibrate the data, the extrapolated prices give us great errors. Hence, although we follow \cite{GatarekBachertMaksymiuk06} and implement the extrapolation, we do not use those two short maturities for the calibration. Moreover, we observed some obvious outliers and corrected them accordingly.

For each of the models and data sets above, we perform three distinct calibrations: first to yields and caplets, then to yields and swaptions, and finally to yields, caplets and swaptions. To define the objective function for each to these calibrations, denote the observed yields and prices of ATM caplets and ATM swaptions by  
$\overline{y}_{0T_i}$, $\overline{\mbox{Cpl}}(T_i,K_{ATM})$ and $\overline{\mbox{Swp}}(T_i,K_{ATM})$, their theoretical counterparties by $y_{0T_i}$, 
$\mbox{Cpl}(T_i,K_{ATM})$ and $\mbox{Swp}(T_i,K_{ATM})$, and the corresponding mean square percentage errors by
\begin{equation}
\text{YieldE}=\sqrt{\frac{1}{n_1}\sum^{n_1}_{i=1}\left[\frac{y_{0T_i}- \overline{y}_{0T_i}}{ \overline{y}_{0T_i}}\right]^2},
\end{equation}
\begin{equation}
\text{CplE}=\sqrt{\frac{1}{n_2}\sum^{n_2}_{i=1}\left[\frac{\text{Cpl}(T_i,K_{ATM})- \overline{\text{Cpl}}(T_i,K_{ATM})}{ \overline{\text{Cpl}}(T_i,K_{ATM})}\right]^2},
\end{equation}
and
\begin{equation}
\text{SwpE}=\sqrt{\frac{1}{n_3}\sum^{n_3}_{i=1}\left[\frac{\text{Swp}(T_i,K_{ATM})- \overline{\text{Swp}}(T_i,K_{ATM})}{ \overline{\text{Swp}}(T_i,K_{ATM})}\right]^2}.
\end{equation}
For the calibration to yields and caplets, we then minimize the objective function
\begin{equation}
\mbox{TotalE1}=\sqrt{\left(\text{YieldE}\right)^2+\left(\text{CplE}\right)^2}.
\end{equation}
Similarly, for the calibration to yields and swaptions, we minimize the objective function
\begin{equation}
\mbox{TotalE2}=\sqrt{\left(\text{YieldE}\right)^2+\left(\text{SwpE}\right)^2}.
\end{equation}
Finally, for the calibration to yields, caplets and swaptions we minimize the objective function 
\begin{equation}
\mbox{TotalE3}=\sqrt{\left(\text{YieldE}\right)^2+\left(\text{CplE}\right)^2+\left(\text{SwpE}\right)^2}.
\end{equation}
For each of these calibrations, we test the pricing performance of the calibrated models. For example, after calibrating to yields and ATM swaptions, we use the model to price the ATM Caplets and compute the pricing error from market ATM Caplet prices. Whenever possible, for example in the Hull-White model, we calibrate initial yield curves and options separately by minimizing their respective square errors, since these involve distinct sets of parameters. Moreover, for the LFM model, we minimize the square error in implied volatilities rather than actual prices.

To access the relative performance between the models, we use the Akaike Information Criterion (AIC) and the model selection relative frequency described in 
\cite{BurnhamAnderson05}. The AIC is formed with the maximized value of the likelihood function $L$ for an estimated model and the number of parameters $k$ in the following way:
\begin{equation}
\text{AIC}=-2\log(L)+2k.\label{002}
\end{equation}
For the least squares method under the assumed normality of residuals, this reduces to .
\begin{equation}
\text{AIC}=n\log\Big(\frac{\text{RSS}}{n}\Big)+2k+n\log(2\pi)+\frac{n}{2},
\end{equation}
where RSS is the fitted residual sum of squares. Since $n\log(2\pi)+\frac{n}{2}$ is a constant, we ignore this term and conclude that
\begin{equation}
\text{AIC}=n\log\Big(\frac{\text{RSS}}{n}\Big)+2k.
\end{equation}
To define the model selection relative frequency, let us suppose we have two models and $N$ calibration sets (for example the different calibration dates in our data). For a data set $j\in\{1,\cdots, N\}$ we compute AIC denoted $AIC^{(1)}_j$ for one model and $AIC^{(2)}_j$ for the other. Suppose the AIC of the first model is smaller than the AIC of the other model $l$ times. Then the model selection relative frequencies (MSRF) for the first model and the second model are computed respectively by 
\begin{equation}
\text{MSRF}^{(1)}=\frac{l}{N} \quad\text{and}\quad
\text{MSRF}^{(2)}=\frac{N-l}{N}.
\end{equation}

\end{document}